\documentstyle[12pt]{article}

\newcommand{\be}{\begin{equation}}
\newcommand{\ee}{\end{equation}}
\newcommand{\Dlt}{\Delta}
\newcommand{\dlt}{\delta}
\newcommand{\prt}{\partial}
\newcommand{\br}{{\bf r}}
\newcommand{\bk}{{\bf k}}
\newcommand{\bj}{{\bf j}}
\newcommand{\bp}{{\bf p}}
\newcommand{\bv}{{\bf v}}
\newcommand{\bt}{\beta}
\newcommand{\vp}{\varphi}
\newcommand{\ep}{\varepsilon}
\newcommand{\al}{\alpha}
\newcommand{\ra}{\rightarrow}
\newcommand{\sgm}{\sigma}
\newcommand{\Sgm}{\Sigma}

\newcommand{\om}{\omega}
\newcommand{\Om}{\Omega}
\newcommand{\Gm}{\Gamma}
\newcommand{\dgr}{\dagger}
\newcommand{\lbd}{\lambda}
\newcommand{\Lbd}{\Lambda}
\newcommand{\cF}{{\cal F}}

\begin{document}

\begin{center}
{\Large{\bf Representative statistical ensembles for Bose systems with
broken gauge symmetry} \\ [5mm]
V.I. Yukalov} \\ [3mm]

{\it Bogolubov Laboratory of Theoretical Physics, \\
Joint Institute for Nuclear Research, Dubna 141980, Russia}

\end{center}

\vskip 2cm

\begin{abstract}

Bose-condensed systems with broken global gauge symmetry are
considered. The description of these systems, as has been shown
by Hohenberg and Martin, possesses an internal inconsistency,
resulting in either nonconserving theories or yielding an unphysical
gap in the spectrum. The general notion of representative statistical
ensembles is formulated for arbitrary statistical systems, equilibrium
or not. The principal idea of this notion is the necessity of taking
into account all imposed conditions that uniquely define the given
statistical system. Employing such a representative ensemble for
Bose-condensed systems removes all paradoxes, yielding a completely
self-consistent theory, both conserving and gapless in any approximation.
This is illustrated for an equilibrium uniform Bose system.

\end{abstract}

\vskip 1cm

{\bf PACS}: 05.30.Ch, 05.30.Jp, 05.70.Ce, 03.75.Hh, 03.75.Kk

\newpage

\section{Introduction and Analysis of Problem}

The appearance of Bose-Einstein condensate in a Bose system is
usually associated with the breaking of the global gauge symmetry,
which is commonly realized by means of the Bogolubov shift in the
field operators [1--4]. The idea of symmetry breaking is in line with
the general understanding that phase transitions of different nature
are accompanied by some symmetry changes. In theoretical description,
there exist several ways of symmetry breaking (see review article [5]).
The most known is the Bogolubov method of quasiaverages, when the
symmetry of the Hamiltonian is disturbed by infinitesimal sources [4].
For breaking the global gauge symmetry, the latter technique is not
always convenient, while the Bogolubov shift is a simple and sufficient 
condition for the symmetry breaking [6].

When the global gauge symmetry is broken, then the description of Bose
systems, based on the standard grand canonical ensemble, encounters a
dilemma, first discussed by Hohenberg and Martin [7], who emphasized that
this description results in either nonconserving theories or yields an
unphysical gap in the spectrum of particles. In nonconserving theories,
local conservation laws are not valid, which, at the same time, is
connected with inconsistencies in thermodynamics.

The Hohenberg-Martin dilemma of conserving versus gapless theories is
usually illustrated by considering some approximate schemes [7], for
instance, a kind of mean-field approximations. The origin of this
dilemma can be explained as follows.

Bose-Einstein condensation implies the macroscopic occupation of
the ground-state energy level, when the number of condensed particles
$N_0$ is such that the condensate fraction $N_0/N$ is nonzero in
the thermodynamic limit. The total number of particles in the system,
$N=N_0+N_1$, becomes a sum of the condensed-particle number $N_0$ and
the number $N_1$ of uncondensed particles. According to Bogolubov [1--4],
the number of condensed particles $N_0$ is such that it provides
thermodynamic stability for the system, minimizing the thermodynamic
potential. At the same time, $N_0=N-N_1$, where $N_1=N_1(\mu,T,\rho)$
is a function of the chemical potential $\mu$, temperature $T$, and
density $\rho\equiv N/V$, with $V$ being the system volume. Hence,
$N_0=N_0(\mu,T,\rho)$ is a function of the same variables. Conversely,
the chemical potential $\mu=\mu(T,\rho)$ is a function of $T$ and $\rho$.
Thus, the condition of the stability, realized as the minimization of
the thermodynamic potential, defines the chemical potential $\mu$. It
is important to emphasize that this procedure of minimization does not
depend on approximations, but is generally valid, as was strictly proved
by Ginibre [8].

From another side, the particle spectrum, under the broken gauge symmetry,
has to be gapless. This is, actually, the condition for the existence of
stable Bose-Einstein condensate. Since, if there would be a gap in the
spectrum, there could be no macroscopic occupation of a single ground-state
level. Hugenholtz and Pines [9] showed that the chemical potential
$\mu=\Sgm_{11}(0,0)-\Sgm_{12}(0,0)$ is expressed through the self-energies
$\Sgm_{\al\bt}(\bk,\om)$, and this relation makes the spectrum gapless for
any Bose system. As far as the self-energies $\Sgm_{\al\bt}(0,0)$ are the
functions of $T$ and $\rho$, the Hugenholtz-Pines relation defines the
chemical potential $\mu=\mu(T,\rho)$.

In this way, the condition of thermodynamic stability, in the frame of the
Bogolubov-Ginibre minimization procedure, defines a chemical potential that
we shall denote as $\mu_{BG}$. At the same time, the existence of a gapless
spectrum, connected to the Hugenholtz-Pines relation, also defines a chemical
potential, which may be denoted as $\mu_{HP}$. These chemical potentials,
found in two different ways do not necessarily coincide, and, generally,
they are different: $\mu_{BG}\neq\mu_{HP}$. Therefore, if one accepts as the
system chemical potential the Bogolubov-Ginibre value $\mu_{BG}$, providing
thermodynamic stability, then one acquires an unphysical gap in the spectrum,
proportional to the difference $|\mu_{HP}-\mu_{BG}|^{1/2}$. Conversely,
accepting as the chemical potential the Hugenholtz-Pines form $\mu_{HP}$,
one gets a nonconserving theory with incorrect thermodynamics, since the
stability condition does not hold. This is the origin of the Hohenberg-Martin
dilemma of conserving versus gapless theories [7].

In any case, the fact that $\mu_{BG}\neq\mu_{HP}$ makes the system unstable
and its description not self-consistent. This problem does not arise only
in the limit of asymptotically weak interactions, when the Bogolubov
weakly-nonideal gas approximation is applicable [1,2]. In this limit,
$\mu_{BG}$ and $\mu_{HP}$ asymptotically coincide. However, in any higher
approximation one has $\mu_{BG}\neq\mu_{HP}$.

For practical applications, one usually does the following. When one
is interested solely in the system dynamics, but not in its spectrum,
one derives the equations from a variational principle, which
guarantees the validity of conservation laws [10,11]. For an
equilibrium system, this is equivalent to the Bogolubov-Ginibre
variational procedure. And, when one studies only the system spectrum,
one accepts the Hugenholtz-Pines relation, forgetting about inconsistent
thermodynamics and instability. Clearly, such palliative ways are not
satisfactory. The principal problem remains how to make the general
theory both conserving as well as gapless.

There have been several attempts to cure the problem, which could be
classified into three groups:

The most often used trick is the omission of anomalous averages. As is
clear, this is a rather unjustified way, since, as soon as the global
gauge symmetry is broken, the anomalous averages do exist and are not
zero. One, sometimes, ascribes this trick to Popov, calling it the "Popov
approximation". However, it is sufficient to look attentively at the original
works by Popov [12--14], which are usually cited in this respect, in order
to realize that he has never suggested anything like that. He considered the
properties of a Bose gas in the vicinity of the critical temperature $T_c$,
honestly calculating all terms, normal and anomalous. When temperature tends
to $T_c$, the condensate density tends to zero together with the anomalous
averages. Actually, both these quantities, the condensate density and the
anomalous averages are the order parameters, appearing together in the
broken-symmetry phase, and disappearing also together, when the symmetry is
getting restored at the critical temperature. Contrary to this, the normal
average, that is, the density of uncondensed particles, increases when
approaching the critical temperature, reaching at $T_c$ the total system
density. This is why at the close vicinity of $T_c$, the anomalous averages
become, without any special assumptions, smaller than the normal ones. However
at low temperatures $T\ll T_c$, the anomalous averages not merely can be of the
order of the normal averages, but can even be much larger than the latter, as
direct calculations show [15]. Moreover, omitting the anomalous averages makes
the system principally unstable [15,16]. Thus, this trick neither has anything
to do with Popov nor can be accepted as a reasonable approximation at low
temperatures. Additionally, if one would wish to ascribe a name to this trick,
one should know that the first person, who really suggested and used it, was
Shohno [17]. And this was known in literature as the Shohno model. The word
"model" is appropriate here, since this is, actually, just a model, but not
a justified approximation. For example, Reatto and Straley [18] used the term
"Shohno model" and studied its properties.

Another way of removing the gap in the spectrum is to calculate the chemical
potential and self energy in different approximations. Then one defines the
chemical potential from the thermodynamic stability condition in one
approximation, but for calculating the self-energy, one invokes a higher-order
approximation, so that to cancel the gap. The additional higher-order terms
can be motivated by Bethe-Salpeter or scattering-matrix approximations [19,20].
This way is what was called by Bogolubov [4] the mismatch of approximations.
Bogolubov already mentioned [4] that such a mismatch can really influence
the appearance or disappearance of the spectrum gap, but renders the theory
not self-consistent and the system unstable.

One can also kill the spectrum gap by adding to the self-energy phenomenological
terms, such that to cancel the gap [21--23]. This way is, clearly, equivalent to
the previous one, since changing the self-energy by invoking some higher-order
approximations is not unique and, hence, is also phenomenological.

As is evident, all these ways, attempting to cure the problem, are based on
a kind of the mismatch of approximations and, as a result, they have the same
common defects:

(i) They are not uniquely defined since there exists an infinite number of
particular tricks for removing the spectrum gap. So, the ambiguity remains
[23].

(ii) They are not self-consistent, involving in this or that way the mismatch
of approximations [4].

(iii) They, as a rule, correspond to an unstable system, either with a not
minimal thermodynamic potential or with a divergent susceptibility [15,16].

(iv) The order of the condensation transitions changes, resulting in a
first-order phase transition, instead of the correct second-order one. This
happens because of the internal inconsistencies in the description. The
disruption of the phase-transition order is a common feature of inconsistent
approximations, which either do not satisfy the stability conditions or
possess a spectrum gap [18,21--23]. This was already noticed in the early
works analysing the Hartree-Fock-Bogolubov approximation having a gap in the
spectrum [18,24--27]. The generality of such a change of the phase-transition
order from second to first in different mean-field approximations was
emphasized in a detailed discussion by Baym and Grinstein [28] and recently
by Kita [23]. As is clear, the thermodynamics of a system with a first-order
phase transition is rather different from that of a system displaying a
second-order transition. And in the vicinity of the critical point, the
thermodynamics in these two cases differs drastically [23,28].

(v) Finally, the mismatch of approximations, by its own, is not a regular
procedure. For each given approximation, it is necessary to invent special
tricks, which, as is stressed above, are not uniquely defined, hence, ambiguous.
There is no a general rule how to do this in a unique way for approximations
of different order.

Thus, we have to conclude that the Hohenberg-Martin dilemma [7] remains
unsolved. The methods, based on the mismatch of approximations, are not
self-consistent, having several internal defects discussed above. In order
that an effective theory be self-consistent, all dynamic and thermodynamic
equations must be derived from the same Hamiltonian, or Lagrangian, and
treated in one chosen approximation, without involving the approximation
mismatch [29,30]. The most recent thorough discussion of the Hohenberg-Martin
dilemma, with many citations, can be found in the review article by
Andersen [31].

In the paper [6], the idea was advanced that the problem can be solved by 
employing a representative statistical ensemble. In Refs. [32,33], it was
shown how to make the Hartree-Fock-Bogolubov (HFB) approximation for a
dilute gas both conserving and gapless. The aim of the present paper is
to develop a general approach, independent of particular approximations,
for the self-consistent treatment of arbitrary Bose systems with broken
gauge symmetry and to demonstrate on the most general footing that the
resulting theory is really completely self-consistent. The approach is
based on the notion of representative statistical ensembles, whose
general formulation is given in Sec. 2 for both equilibrium as well
as nonequilibrium systems. This notion is specified in Sec. 3 for
Bose systems with broken gauge symmetry. Thermodynamic self-consistency
of the theory is emphasized in Sec. 4. Operator equations of motion are
obtained in Sec. 5 for a Hamiltonian with an arbitrary interaction
potential. Sec. 6 demonstrates that the local conservation laws are
valid on the operator level, hence, being automatically satisfied for 
the related average quantities. The equation for the condensate wave 
function is derived and analyzed in Sec. 7 for an arbitrary Bose system. 
In Sec. 8 a uniform Bose system is considered and illustrated for the HFB
approximation. The behaviour of the condensate and superfluid fractions
is studied in Sec. 9. The equations for Green functions are presented in
Sec. 10, where it is shown, by using the Bogolubov theorem [4], that the
Hugenholtz-Pines relation follows, thus, proving the complete
self-consistency of the developed approach. Sec. 11 is the conclusion.

Throughout the paper, the system of units is used, where the Planck and
Boltzmann constants are set to unity, $\hbar=1$, $k_B=1$.

\section{Representative Statistical Ensembles}

The idea of representative statistical ensembles goes back to Gibbs [34],
who mentioned that to prescribe a canonical distribution for a system
may be not sufficient, but this distribution has to be complimented by
those constraints and conditions that provide a correct representation
of the considered statistical system. The term "representative ensembles"
was used by ter Haar [35,36], who investigated the problem of a proper
representation of equilibrium statistical systems. Equilibrium and
quasiequilibrium representative ensembles were described in the
review article [5]. In this section, we shall formulate the notion
of representative statistical ensembles for both equilibrium and
nonequilibrium systems.

Representative statistical ensembles for equilibrium systems are sometimes
also termed as generalized Gibbs ensembles, subjective or conditional ensembles.
Their mathematical construction is based on the conditional maximization of the
Gibbs entropy, as was done by Janes [37,38]. This concept was also employed 
by Girardeau [39].

Let us start with equilibrium systems. An {\it equilibrium statistical
ensemble}, by definition, is a pair $\{\cF,\hat\rho\}$ of the space $\cF$
of microstates and a statistical operator $\hat\rho$. In order to correctly
define the latter, it is necessary, according to Gibbs [34], to take into
account all conditions and constraints, imposed on the system. Suppose, we
have a set $\{\hat C_i\}$ of self-adjoint operators, defined on the space
$\cF$, which will be called condition operators. This is because these
operators enter the {\it statistical conditions}
\be
\label{1}
C_i = \; <\hat C_i>\; = \; {\rm Tr}\hat\rho\; \hat C_i \; ,
\ee
which are necessary to take into account for correctly representing the
considered system. The trace operation in Eq. (1) is over the given space
$\cF$, which can be defined as a Fock space. The first evident constraint
is the normalization condition for the statistical operator,
\be
\label{2}
1 \; = \; < \hat 1_\cF> \; =  \; {\rm Tr}\hat\rho \; ,
\ee
where $\hat 1_\cF$ is the unity operator in $\cF$. The definition of the
internal energy
\be
\label{3}
E\; = \; <\hat H> \; = \; {\rm Tr}\hat\rho\; \hat H \; ,
\ee
as the average of a Hamiltonian $\hat H$, is another standard statistical
condition. But, in addition to constraints (2) and (3), there can exist any
number of other statistical conditions (1). The conditional maximization of
the Gibbs entropy
\be
\label{4}
S \equiv - {\rm Tr}\hat\rho \; \ln\hat\rho
\ee
is equivalent to the unconditional minimization of the {\it information
functional} [40], defined as
\be
\label{5}
I[\hat\rho] = - S + \lbd_0\left ({\rm Tr}\hat\rho - 1\right ) +
\bt \left ({\rm Tr}\hat\rho\; \hat H - E \right ) + \bt \sum_i \nu_i
\left ({\rm Tr}\hat\rho\; \hat C_i - C_i \right ) \; ,
\ee
in which the standard conditions (2) and (3) are included explicitly.
The quantities $\lbd_0$, $\bt$, and $\bt\nu_i$ are the appropriate
Lagrange multipliers, with $\bt\equiv 1/T$ being inverse temperature.
Minimizing the information functional (5) with respect to the statistical
operator $\hat\rho$ gives
\be
\label{6}
\hat\rho = \frac{1}{Z} \; e^{-\bt H} \; ,
\ee
where $Z\equiv\exp(1+\lbd_0)$ is the partition function and the {\it
grand Hamiltonian}
\be
\label{7}
H \equiv \hat H + \sum_i \nu_i \hat C_i
\ee
is introduced. The representative statistical ensemble, under
constraints (1), (2), and (3), is then the pair $\{\cF,\hat\rho\}$
of a space $\cF$ of microstates and the statistical operator (6),
with the grand Hamiltonian (7). When one of the condition operators
$\hat C_i$ is the number-of-particle operator $\hat N$ and the related
Lagrange multiplier $\nu_i=-\mu$, one gets a particular form of the
grand Hamiltonian $H=\hat H-\mu\hat N$. However, any other necessary
constraints can be included, resulting in the general expression (7)
for the grand Hamiltonian.

The condition operators $\hat C_i$ are to be self-adjoint,
$\hat C_i^+=\hat C_i$, so that the grand Hamiltonian (7) be also
self-adjoint. In many cases, the condition operators are taken as
integrals of motion, such that $[\hat C_i,\hat H]=0$. But this is not
compulsory. For instance, the number-of-particle operator $\hat N$
does not commute with the Hamiltonian energy operator $\hat H$, when
the global gauge symmetry is broken.

The representative ensemble $\{\cF,\hat\rho\}$, with the statistical
operator (6) and the grand Hamiltonian (7), define all thermodynamics
of an equilibrium, or stationary, system. The construction of this
ensemble has been more or less straightforward, following the ideas of
Gibbs [34], ter Haar [35,36], and Janes [37,38], as is reviewed in Refs.
[5,40]. But the definition of representative ensembles for arbitrary
nonequilibrium systems is not evident. Below, we give the generalization
of the notion of representative ensembles for nonequilibrium statistical
systems.

To describe a nonequilibrium system, one needs, in addition to the space
of microstates $\cF$ and the initial value of the statistical operator
$\hat\rho=\hat\rho(0)$, to define the temporal evolution of the system.
This evolution can be described by the time-dependent statistical
operators $\hat\rho(t)$, satisfying the Liouville equation, or by the
time dependence of physical operators, satisfying the Heisenberg equation.
Equivalently, the time evolution can be associated with the evolution
operator, satisfying the Schr\"odinger equation. Keeping in mind any of
these ways, we may denote the prescribed temporal evolution by the symbol
$\prt/\prt t$. Then a {\it nonequilibrium statistical ensemble} is a
triplet $\{\cF,\hat\rho,\prt/\prt t\}$. Clearly, when the time evolution
is absent, or trivial, this definition reduces to that for the equilibrium
case.

To be more specific, we may remember that each system is characterized by
some dynamical variables, such as field operators. Let us keep in mind a
set of field operators $\psi(x,t)$, whose particular representation is not
important at this stage. For example, the variable $x$ can represent
real-space coordinates or momenta. It may also include other continuous
or discrete variables, such as the spin indices or component labels. All
physical operators, such as the Hamiltonian energy operator $\hat H[\psi]$,
are functionals of the field operators.

The most general way for describing the system dynamics, as is known [41],
is the extremization of the action functional. Implementing this for our
case, we need the Lagrangian
\be
\label{8}
\hat L[\psi] \equiv \hat E[\psi] - \hat H[\psi] \; ,
\ee
in which the notation
\be
\label{9}
\hat E[\psi] \equiv \int \psi^\dgr(x,t)\; i \;
\frac{\prt}{\prt t}\; \psi(x,t)\; dx
\ee
for the temporal energy operator is used.

To make the ensemble representative, we have to take account of all
additional conditions uniquely characterizing the system. This implies
that, similarly to Eq. (1), we have to take care of the {\it statistical
conditions}
\be
\label{10}
C_i \; = \; <\hat C_i[\psi]> \; ,
\ee
where $\hat C_i[\psi]$ are the appropriate condition operators and
$$
<\hat C_i[\psi]> \; \equiv \; {\rm Tr}\hat\rho(0)\;
\hat C_i[\psi(x,t)] = {\rm Tr}\hat\rho(t)\; \hat C_i [\psi(x,0)] \; .
$$

The principle of action extremization, under the given statistical
conditions (10), is equivalent to the unconditional extremization of
the effective action
\be
\label{11}
A[\psi] = \int \left\{ \hat L[\psi] -\nu_i \hat C_i[\psi]
\right \}\; dt
\ee
with the Lagrange multipliers $\nu_i$ guaranteeing the validity
of conditions (10). Combining Eqs. (8) and (11), we can rewrite the
effective action (11) as
\be
\label{12}
A[\psi] = \int \left\{ \hat E[\psi] - H[\psi]\right \} \; dt \; ,
\ee
with the {\it grand Hamiltonian}
\be
\label{13}
H[\psi] \equiv \hat H[\psi] + \sum_i \nu_i \; \hat C_i[\psi] \; ,
\ee
having the same form as in Eq. (7). The extremization of the action
functional with respect to field operators means the variational
equation
\be
\label{14}
\frac{\dlt A[\psi]}{\dlt\psi^\dgr(x,t)} = 0
\ee
plus its Hermitian conjugate. Equation (14), in view of action (12),
is identical to the equation
\be
\label{15}
i\; \frac{\prt}{\prt t}\; \psi(x,t) =
\frac{\dlt H[\psi]}{\dlt\psi^\dgr(x,t)} \; .
\ee

The evolution equation (15) is what one needs for a complete
definition of the nonequilibrium representative ensemble. It is
important to stress that the system dynamics is governed by the same
grand Hamiltonian as its thermodynamics.

We may also note that in the Heisenberg representation, as is well
known (see, e.g., Refs. [40,41]), the variational equation (15) is
the same as the Heisenberg equation
$$
i\; \frac{\prt}{\prt t}\; \psi(x,t) =
\left [ \psi(x,t), \; H[\psi] \right ] \; .
$$
The time evolution of the field operators is given by the form
$$
\psi(x,t) = \hat U^+(t) \psi(x,0) \hat U(t) \; ,
$$
expressed through the evolution operator $\hat U(t)$ satisfying the
Schr\"odinger equation
$$
i\; \frac{d}{dt} \; \hat U(t) =  H[\psi(x,0)] \hat U(t) \; .
$$
Respectively, the time dependence of the statistical operator
$\hat\rho(t)$ stems from the Liouville equation, yielding
$$
\hat\rho(t) = \hat U(t)\; \hat\rho(0)\; \hat U^+(t) \; .
$$
In any case, it is the grand Hamiltonian (13), which governs the temporal
evolution of a nonequilibrium system.

\section{Broken Gauge Symmetry}

Now, let us specify the general notion of representative statistical
ensembles, formulated above, for Bose systems, in which there exists the
critical temperature $T_c$ below which the global $U(1)$ gauge symmetry
becomes broken. For concreteness, we keep in mind a one-component system
of spinless particles, characterized by the field operators satisfying
the Bose commutation relations.

Above the critical temperature $T_c$, the system is described by field
operators $\psi=\psi(\br,t)$ and the conjugate $\psi^\dgr$, being
functions of spatial, $\br$, and temporal, $t$, variables. These
operators generate the Fock space $\cF(\psi)$ on which all physical
operators are defined. The related mathematical details can be found in
the books [40--42]. The Hamiltonian energy operator $\hat H[\psi]$ is a
gauge-invariant functional of the field operators. The number-of-particle
operator $\hat N[\psi]$ is normalized to the total number of particles
$N=<\hat N[\psi]>$. Therefore the grand Hamiltonian is
$$
H[\psi] = \hat H[\psi] - \mu \hat N[\psi] \qquad (T > T_c) \; .
$$
This grand Hamiltonian, entering the statistical operator (6),
characterizes the representative statistical ensemble for the normal Bose
system, above the critical temperature, where the global gauge symmetry is
preserved.

Below the critical temperature $T_c$, the global gauge symmetry becomes
broken. The symmetry breaking is realized by the Bogolubov shift
\be
\label{16}
\psi(\br,t) \longrightarrow \; \hat \psi(\br,t) \; \equiv
\; \eta(\br,t) + \psi_1(\br,t) \; ,
\ee
where $\eta(\br,t)$ is the condensate wave function, while $\psi_1(\br,t)$
is the field operator of uncondensed particles. The field operators $\psi_1$
and $\psi_1^\dgr$ generate the Fock space $\cF(\psi_1)$, which all physical
operators are to be defined on. In Eq. (16), the condensate wave function
$\eta(\br,t)$, strictly speaking, is assumed to be factored with the unity
operator $\hat 1_\cF$ in $\cF(\psi_1)$. However here and in what follows,
we shall use the common way of omitting the explicit appearance of the unity
operator, in order not to make formulas too cumbersome. The condensate
function $\eta(\br,t)$ can also be termed as the coherent field, since the
related coherent state $|\eta>$ is the vacuum state in the space $\cF(\psi_1)$.
This and other mathematical details can be found in Refs. [40,42,43].

Thus, instead of one field variable $\psi(\br,t)$ above $T_c$, for the
Bose system below $T_c$, where the gauge symmetry is broken, there arise
two field variables, the condensate function (coherent field) $\eta(\br,t)$
and the field operator $\psi_1(\br,t)$ of uncondensed particles. These two
dynamical variables are, of course, assumed to be linearly independent,
being orthogonal to each other,
\be
\label{17}
\int \eta^*(\br,t) \psi_1(\br,t)\; d\br = 0 \; .
\ee

It is of principal importance to emphasize that the spaces $\cF(\psi)$
for $T>T_c$ and $\cF(\psi_1)$ for $T<T_c$ are mutually orthogonal [43,44].
The field operators $\psi$ and $\psi_1$ are defined on different spaces,
$\cF(\psi)$ and $\cF(\psi_1)$, respectively, realizing two different
unitary nonequivalent operator representations, with the Bose commutation
relations [43,44]. As soon as the gauge symmetry is broken, one has to deal
with the space $\cF(\psi_1)$. It would be principally incorrect to work,
first, in the space $\cF(\psi)$, accomplishing there some transformations,
and then to pass to the space $\cF(\psi_1)$ by breaking the symmetry with
the Bogolubov shift (16). As is shown in Ref. [6], such a procedure leads to
wrong results. From the mathematical point of view, it is absolutely obvious
that any manipulations must be accomplished in one given space, where all
operators are defined.

Now, in the space $\cF(\psi_1)$, we have two normalization conditions for
two linearly independent field variables, $\eta(\br,t)$ and $\psi_1(\br,t)$.
The condensate function is normalized to the number of condensed particles
\be
\label{18}
N_0 = \int |\eta(\br,t)|^2 \; d\br \; ,
\ee
which is assumed to be macroscopic, such that the condensate fraction
$N_0/N$ be nonzero in the thermodynamic limit. This normalization can be
rewritten in the standard form of the statistical conditions (10) by using
the operator
\be
\label{19}
\hat N_0 \equiv N_0 \hat 1_\cF \; ,
\ee
where $\hat 1_\cF$ is the unity operator in $\cF(\psi_1)$. Then Eq. (18)
transforms to the statistical condition
\be
\label{20}
N_0 \; =  \; < \hat N_0> \; ,
\ee
in which, as in what follows, the averaging is over the space
$\cF(\psi_1)$. The second normalization is, clearly, for the number of
uncondensed particles
\be
\label{21}
N_1 \; = \; < \hat N_1> \; ,
\ee
with the corresponding operator
\be
\label{22}
\hat N_1 \; \equiv \; \int \psi_1^\dgr(\br,t)
\psi_1(\br,t)\; d\br \; .
\ee
The total number of particles
\be
\label{23}
N \; = \; <\hat N> \; = \; N_0 + N_1
\ee
is the average of the operator
\be
\label{24}
\hat N \; \equiv \; \int \hat\psi^\dgr(\br,t)
\hat\psi(\br,t)\; d\br \; = \; \hat N_0 + \hat N_1 \; .
\ee
Normalization (23) follows from Eqs. (20) and (21). Therefore, among three
normalization conditions, (20), (21), and (23), only two can be treated as
independent. Generally, any combination of the pairs, for $N_0$ and $N_1$,
or for $N_0$ and $N$, or for $N_1$ and $N$, could be chosen. For the sake
of symmetry, we prefer to choose the normalization conditions (20) and (21).

When the gauge symmetry is broken, the average $<\psi_1>$ may become nonzero.
This, however, would mean that quantum numbers, as spin or momentum, are not
conserved. In order to avoid this unpleasant situation, one has to impose an
additional constraint
\be
\label{25}
<\psi_1(\br,t)> \; = \; 0 \; .
\ee
The latter can be reduced to the standard form of the statistical conditions
(10) by defining a self-adjoint operator
\be
\label{26}
\hat\Lbd[\psi_1] \; \equiv \;
\int \left [ \lbd(\br,t) \psi_1^\dgr(\br,t) + \lbd^*(\br,t)\psi_1(\br,t)
\right ]\; d\br \; ,
\ee
in which $\lbd(\br,t)$ is a complex function. Then constraint (25) can be
rewritten as the {\it quantum-number conservation condition}
\be
\label{27}
<\hat\Lbd[\psi_1]> \; = \; 0 \; .
\ee

In this way, for the correct representation of a Bose system with
broken gauge symmetry, we must work in the space $\cF(\psi_1)$ and take
into account three statistical conditions, (20), (21), and (27). The
corresponding representative ensemble is constructed following the general
procedure, formulated in Sec. 2.

For an equilibrium system, according to Eq. (5), we have the information
functional
$$
I[\hat\rho] ={\rm Tr}\hat\rho\; \ln\hat\rho + \lbd_0 \left (
{\rm Tr}\hat\rho -1 \right ) + \bt \left ( {\rm Tr}\hat\rho\;
\hat H[\eta,\psi_1] - E \right ) 
$$
\be
\label{28}
 - \; \bt\mu_0 \left ( {\rm Tr}\hat\rho \; \hat N_0 - N_0 \right ) -
\bt\mu_1 \left ( {\rm Tr}\hat\rho \; \hat N_1 - N_1 \right ) -
\bt {\rm Tr}\hat\rho \; \hat \Lbd[\psi_1] \; .
\ee
Minimizing the latter yields the statistical operator
\be
\label{29}
\hat\rho = \frac{1}{Z} \; \exp\left\{ -\bt H[\eta,\psi_1]
\right \} \; ,
\ee
with the partition function
$$
Z \equiv {\rm Tr} \exp\left\{ -\bt H[\eta,\psi_1]
\right \} \; ,
$$
and the grand Hamiltonian
\be
\label{30}
H[\eta,\psi_1] \equiv \hat H [\eta,\psi_1] -\mu_0 \hat N_0 -
\mu_1\hat N_1 -\hat\Lbd[\psi_1] \; ,
\ee
in agreement with Eq. (7).

For the general case of a nonequilibrium system, we have the Lagrangian
$$
\hat L [\eta,\psi_1] = \hat E[\eta,\psi_1] - \hat H[\eta,\psi_1] \; ,
$$
in which
$$
\hat E[\eta,\psi_1] =\int \left\{ \eta^*(\br,t)\; i \;
\frac{\prt}{\prt t}\; \eta(\br,t) + \psi_1^\dgr(\br,t)\; i \;
\frac{\prt}{\prt t} \; \psi_1(\br,t)\right \} \; d\br \; .
$$
Then, similarly to Eq. (12), the effective action is
\be
\label{31}
A[\eta,\psi_1] = \int \left\{ \hat E[\eta,\psi_1] - H][\eta,\psi_1]
\right \} \; dt \; ,
\ee
with the same grand Hamiltonian (30). Since we have now two linearly
independent field variables, the extremization of action (31) gives
two variational equations, for the condensate function,
\be
\label{32}
\frac{\dlt A[\eta,\psi_1]}{\dlt\eta^*(\br,t)} = 0 \; ,
\ee
and for the field operators of uncondensed particles,
\be
\label{33}
\frac{\dlt A[\eta,\psi_1]}{\dlt\psi^\dgr_1(\br,t)} = 0 \; .
\ee
Equations (32) and (33), in view of the effective action (31), are
equivalent to the evolution equations
\be
\label{34}
i\; \frac{\prt}{\prt t}\; \eta(\br,t) =
\frac{\dlt H[\eta,\psi_1]}{\dlt\eta^*(\br,t)}
\ee
and, respectively,
\be
\label{35}
i\; \frac{\prt}{\prt t}\; \psi_1(\br,t) =
\frac{\dlt H[\eta,\psi_1]}{\dlt\psi_1^\dgr(\br,t)} \; .
\ee
And, as usual, these equations are to be complimented by their
Hermitian conjugate.

Thus, the representative statistical ensemble
$\{\cF(\psi_1),\hat\rho,\prt/\prt t\}$ for a Bose system with broken
global gauge symmetry is defined in complete agreement with the general
theory of Sec. 2. The dynamics and thermodynamics of such a system are
governed by the grand Hamiltonian (30). It is only by accurately taking
into account all imposed constraints that it is possible to correctly
define the representative ensemble and to develop a self-consistent
theory, avoiding any internal defects and paradoxes. The imposed
statistical conditions (20), (21), and (27) lead to the grand Hamiltonian
(30), with the Lagrange multipliers $\mu_0$, $\mu_1$, and $\lbd(\br,t)$.
The form of this Hamiltonian is more general than that of the trivial
expression $\hat H-\mu\hat N$. For a system with broken gauge symmetry, 
the number of the Lagrange multipliers in the form $\hat H -\mu\hat N$ is 
smaller than the number of imposed constraints. As a result, the problem 
becomes mathematically overdefined, which leads to the inconsistencies 
described in Sec. 1.

\section{Self-Consistent Thermodynamic Relations}

For an equilibrium system, the condensate function does not depend on
time, $\eta(\br,t)=\eta(\br)$. The grand thermodynamic potential
\be
\label{36}
\Om = - T\ln{\rm Tr}\exp\left ( -\bt H[\eta,\psi_1] \right )
\ee
is defined through the grand Hamiltonian (30). As is evident by this
definition,
$$
\frac{\prt\Om}{\prt\mu_0} = -N_0 \; , \qquad
\frac{\prt\Om}{\prt\mu_1} = -N_1 \; .
$$
Varying potential (36) with respect to $\eta(\br)$ gives the equation
\be
\label{37}
\frac{\dlt\Om}{\dlt\eta^*(\br)} = \;
< \frac{\dlt H[\eta,\psi_1]}{\dlt\eta^*(\br)} > \; = \; 0
\ee
for the condensate function, in agreement with Eq. (34), when in
equilibrium $\prt\eta(\br)/\prt t=0$. In order that the system with
Bose-Einstein condensate be stable, the thermodynamic potential (36)
is to be minimal with respect to the number of condensed particles $N_0$,
so that
\be
\label{38}
\frac{\dlt\Om}{\dlt N_0} = \;
< \frac{\prt H[\eta,\psi_1]}{\prt N_0} > \; = \; 0 \; ,
\ee
which is the Bogolubov-Ginibre stability condition. Under normalization
(18), the dependence on $N_0$ comes through the condensate function
$\eta(\br)$, because of which
$$
\frac{\prt\Om}{\prt N_0} = \int \left [
\frac{\dlt\Om}{\dlt\eta(\br)}\; \frac{\prt\eta(\br)}{\prt N_0}
+ \frac{\dlt\Om}{\dlt\eta^*(\br)}\; \frac{\prt\eta^*(\br)}{\prt N_0}
\right ]\; d\br \; .
$$
Therefore Eq. (38) is a direct consequence of Eq. (37). Of course, for
a uniform system, Eqs. (37) and (38) identically coincide.

The average densities of condensed, uncondensed, and all particles are
\be
\label{39}
\rho_0 \equiv \frac{N_0}{V} \; , \qquad \rho_1 \equiv \frac{N_1}{V} \; ,
\qquad \rho \equiv \frac{N}{V} \; ,
\ee
respectively. The related fractions of condensed and uncondensed particles
are
\be
\label{40}
n_0 \equiv \frac{N_0}{N} \equiv \frac{\rho_0}{\rho} \; , \qquad
n_1 \equiv \frac{N_1}{N} \equiv \frac{\rho_1}{\rho} \; .
\ee
From these definitions, one has
\be
\label{41}
\rho = \rho_0 + \rho_1 \; , \qquad n_0 + n_1 =  1 \; .
\ee
What one can usually fix in experiment is temperature $T$ and the total
density $\rho$.

The system free energy  writes as
\be
\label{42}
F = \Om + \mu_0 N_0 + \mu_1 N_1 \; .
\ee
Recall that $N_0 = N_0(T,\rho)$ is such that to guarantee the system
thermodynamic stability. From the equations of motion, together with the
relation $N=N_0+N_1$, one finds $N_1=N_1(T,\rho)$. As is mentioned above,
in experiment, only the total number of particles can be fixed, except
temperature and volume. Hence, the free energy can be written as
\be
\label{43}
F = \Om + \mu N \; ,
\ee
which can be considered as the definition of the system chemical
potential $\mu$. Comparing Eqs. (42) and (43) gives
\be
\label{44}
\mu=\mu_0 n_0 +  \mu_1 n_1 \; .
\ee
Then one can define the free energy $F=F(T,V,N)$ as a function of temperature,
volume, and the total number of particles, with the differential
\be
\label{45}
dF = -S\; dT - P\; dV + \mu\; dN \; ,
\ee
in which $S$ is entropy and $P$ is pressure. From here, all measurable
thermodynamic quantities are calculated in the standard way. For example,
the system chemical potential is
\be
\label{46}
\mu = \left ( \frac{\prt F}{\prt N} \right )_{TV} \; .
\ee
As is evident, Eqs. (44) and (46) are identical by the above definitions.

At zero temperature, the free energy $F=E-TS$ reduces to the internal
energy
\be
\label{47}
E \equiv \; < \hat H[\eta,\psi_1]> \; = \; < H[\eta,\psi_1]> + \mu N \; .
\ee
Therefore the chemical potential (46) becomes
$$
\mu = \left ( \frac{\prt E}{\prt N} \right )_V \qquad
(T=0) \; .
$$

The grand potential (36) is a function $\Om=\Om(T,V,\mu)$ of temperature,
volume, and chemical potential, with the differential
\be
\label{48}
d\Om = - S\; dT - P\; dV - N\; d\mu \; .
\ee
It is easy to check that the thermodynamic and Gibbs entropies coincide,
so that
\be
\label{49}
S = - \; \frac{\prt\Om}{\prt T} = - {\rm Tr}\hat\rho \;
\ln\hat\rho \; .
\ee
This immediately follows from the direct differentiation of the grand
potential (36) and the form of the statistical operator (29).

It is also possible to prove (see the proof in the Appendix A), that the
standard relation
\be
\label{50}
\frac{\prt N}{\prt\mu} = \bt\Dlt^2(\hat N)
\ee
holds, in which $\Dlt^2(\hat N)$ is the dispersion of the number-of-particle
operator $\hat N=\hat N_0+\hat N_1$. The dispersion of a self-adjoint operator
$\hat A$ is
$$
\Dlt^2(\hat A) \equiv \; < \hat A^2> - <\hat A>^2 \; .
$$

In this way, all thermodynamics quantities and relations are defined
self-consistently.

\section{Operator Equations of Motion}

The equations of motion for the field variables $\eta(\br,t)$ and
$\psi_1(\br,t)$ are given by Eqs. (34) and (35). To specify them, let
us take the Hamiltonian energy operator in the standard form
$$
\hat H[\eta,\psi_1] = \int \hat\psi^\dgr(\br) \left ( - \;
\frac{\nabla^2}{2m} + U \right ) \hat\psi(\br) \; d\br 
$$
\be
\label{51}
+ \; \frac{1}{2}\; \int \hat\psi^\dgr(\br) \hat\psi^\dgr(\br')
\Phi(\br-\br') \hat\psi(\br')\hat\psi(\br) \; d\br d\br' \; ,
\ee
in which $\hat\psi(\br)=\hat\psi(\br,t)$ is the shifted field operator
defined in Eq. (16). In order to avoid cumbersome notations, we omit
the explicit dependence on time of the variables $\eta(\br)=\eta(\br,t)$
and $\psi_1(\br)=\psi_1(\br,t)$, though this dependence is assumed. In
what follows, such an abreviated notation will be often used, where it
does not lead to confusion. The external potential $U=U(\br,t)$ can,
in general, depend on time. The interaction potential $\Phi(\br)$ is
arbitrary, provided it is symmetric, such that $\Phi(\br)=\Phi(-\br)$,
and enjoys the Fourier transformation.

The grand Hamiltonian (30) can be written as a sum of five terms,
\be
\label{52}
H[\eta,\psi_1] = \sum_{n=0}^4 H^{(n)} \; ,
\ee
whose order is labelled by the number of operators $\psi_1$ in their
products. The zero-order term contains no $\psi_1$, being
$$
H^{(0)} = \int \eta^*(\br) \left ( -\; \frac{\nabla^2}{2m} + U -
\mu_0 \right ) \eta(\br) \; d\br 
$$
\be
\label{53}
+ \; \frac{1}{2} \int \Phi(\br-\br') |\eta(\br')|^2 |\eta(\br)|^2 \;
d\br d\br' \; .
\ee
To satisfy the quantum-number conservation condition (27), the Hamiltonian
must not contain the terms linear in $\psi_1$. This prescribes to take the
Lagrange multipliers in Eq. (26) as
\be
\label{54}
\lbd(\br,t) = \left [ - \; \frac{\nabla^2}{2m} + U +
\int \Phi(\br-\br') |\eta(\br',t)|^2\; d\br' \right ] \eta(\br,t) \; ,
\ee
so that $H^{(1)}=0$. The necessity of removing the terms linear in $\psi_1$
or $\psi_1^\dgr$, by means of condition (54), in order to satisfy constraint
(25), can be easily proved for quadratic Hamiltonians, involving linear
terms, by employing the method of canonical transformations [42,45]. In the
general case of arbitrary Hamiltonians, the proof of Eq. (54), yielding the
cancellation of linear terms, is presented in the Appendix B.

The second-order term is
$$
H^{(2)} = \int \psi_1^\dgr(\br) \left ( - \; \frac{\nabla^2}{2m} +
U - \mu_1 \right ) \psi_1(\br)\; d\br 
$$
$$
+ \; \int \Phi(\br-\br') \left [
|\eta(\br)|^2 \psi_1^\dgr(\br') \psi_1(\br') +
\eta^*(\br)\eta(\br')\psi_1^\dgr(\br')\psi_1(\br) \right.
$$
\be
\label{55}
+ \; \left. \frac{1}{2}\; \eta^*(\br) \eta^*(\br') 
\psi_1(\br')\psi_1(\br) + \frac{1}{2}\; \eta(\br)\eta(\br') 
\psi_1^\dgr(\br') \psi_1^\dgr(\br) \right ] \; d\br d\br' \; .
\ee
Respectively, we have the third-order term
\be
\label{56}
H^{(3)} = \int \Phi(\br-\br') \left [
\eta^*(\br)\psi_1^\dgr(\br')\psi_1(\br')\psi_1(\br) +
\psi_1^\dgr(\br) \psi_1^\dgr(\br') \psi_1(\br') \eta(\br) \right ]\;
d\br d\br'
\ee
and the fourth-order term
\be
\label{57}
H^{(4)} = \frac{1}{2} \; \int \psi_1^\dgr(\br) \psi_1^\dgr(\br')
\Phi(\br-\br') \psi_1(\br')\psi_1(\br)\; d\br d\br' \; .
\ee

From Eq. (34), we obtain
\be
\label{58}
i\; \frac{\prt}{\prt t} \; \eta(\br,t) = \left ( - \;
\frac{\nabla^2}{2m} + U - \mu_0 \right ) \eta(\br,t) +
\int \Phi(\br-\br') \left [ |\eta(\br')|^2 \eta(\br) +
\hat X(\br,\br') \right ] \; d\br' \; ,
\ee
where the last term in the integrand is the correlation operator
$$
\hat X(\br,\br') \equiv \psi_1^\dgr(\br')\psi_1(\br')\eta(\br) 
$$
\be
\label{59}
+ \; \psi_1^\dgr(\br')\eta(\br')\psi_1(\br) +
\eta^*(\br')\psi_1(\br')\psi_1(\br) +
\psi_1^\dgr(\br')\psi_1(\br')\psi_1(\br) \; .
\ee
And Eq. (35) yields
$$
i\; \frac{\prt}{\prt t}\; \psi_1(\br,t) = \left ( - \;
\frac{\nabla^2}{2m} + U - \mu_1 \right )\psi_1(\br,t) 
$$
\be
\label{60}
+\; \int \Phi(\br-\br') \left [ |\eta(\br')|^2 \psi_1(\br) +
\eta^*(\br')\eta(\br)\psi_1(\br') +
\eta(\br')\eta(\br)\psi_1^\dgr(\br') + \hat X(\br,\br')\right ]\;
d\br' \; .
\ee
Again, for brevity, the time dependence is not explicitly shown in the
integrals of Eqs. (59) and (60). Also, recall that in the operator equation
(58), the condensate function is assumed to be factored with $\hat 1_\cF$,
the unity operator in $\cF(\psi_1)$.

\section{Local Conservation Laws}

The equations of motion (58) and (60) are derived from the variational
principle of action extremization. Therefore they guarantee the validity
of all local conservation laws on the operator level. As an illustration,
let us consider the time variation of the local densities.

The local condensate density is
\be
\label{61}
\rho_0(\br,t) \equiv |\eta(\br,t)|^2 \; ,
\ee
and the local condensate current is
\be
\label{62}
\bj_0(\br,t) \equiv -\; \frac{i}{2m} \left [ \eta^*(\br,t)\;
\nabla\eta(\br,t) - \eta(\br,t)\; \nabla\eta^*(\br,t)\right ] \; .
\ee
Respectively, we define the operator density of uncondensed particles
\be
\label{63}
\hat\rho_1(\br,t) \equiv \psi_1^\dgr(\br,t)\psi_1(\br,t)
\ee
and the related current operator
\be
\label{64}
\bj_1(\br,t) \equiv -\; \frac{i}{2m} \left\{ \psi_1^\dgr(\br,t)\;
\nabla\psi_1(\br,t) - \left [\nabla\psi_1^\dgr(\br,t)\right ]
\psi_1(\br,t)\right \} \; .
\ee

Differentiating the condensate density (61), we find
\be
\label{65}
\frac{\prt}{\prt t}\; \rho_0(\br,t) + \nabla\cdot\bj_0(\br,t) =
- \hat \Gm(\br,t) \; ,
\ee
where the source operator is
\be
\label{66}
\hat\Gm(\br,t) \equiv \int \Phi(\br-\br') \left [
\hat R(\br,\br') + \hat R^+(\br,\br')\right ]\; d\br' \; ,
\ee
in which
$$
\hat R(\br,\br') \equiv i\eta^*(\br) \left [
\psi_1^\dgr(\br')\eta(\br') + \eta^*(\br')\psi_1(\br') +
\psi_1^\dgr(\br')\psi_1(\br')\right ] \psi_1(\br) \; .
$$
And the time variation of density (63) yields
\be
\label{67}
\frac{\prt}{\prt t} \; \hat\rho_1(\br,t) +
\nabla\cdot\hat\bj_1(\br,t) = \hat\Gm(\br,t) \; .
\ee
For the total operator density
\be
\label{68}
\hat\rho(\br,t) = \rho_0(\br,t) + \hat\rho_1(\br,t)
\ee
and the total operator of current
\be
\label{69}
\hat\bj(\br,t) = \bj_0(\br,t) + \hat\bj_1(\br,t) \; ,
\ee
we obtain the {\it continuity equation}
\be
\label{70}
\frac{\prt}{\prt t}\; \rho(\br,t) +
\nabla\cdot \hat\bj(\br,t) = 0 \; .
\ee
In the same way, one can derive any other local conservation laws, following
the standard procedure [40,45] and employing the equations of motion (58)
and (60).

\section{Condensate Wave Function}

The equation for the condensate wave function $\eta(\br,t)$ follows from
averaging Eq. (58) over $\cF(\psi_1)$. To this end, we need to introduce
several notations. The normal density matrix is
\be
\label{71}
\rho_1(\br,\br') \; \equiv \; < \psi_1^\dgr(\br')\psi_1(\br)> \; .
\ee
Under the broken gauge symmetry, there appears the so-called anomalous
density matrix
\be
\label{72}
\sgm_1(\br,\br') \; \equiv \; <\psi_1(\br') \psi_1(\br)> \; .
\ee
The density of condensed particles is $\rho_0(\br)$, defined in Eq. (61),
and the density of uncondensed particles is
\be
\label{73}
\rho_1(\br) \equiv \rho_1(\br,\br) \; = \;
<\psi_1^\dgr(\br)\psi_1(\br)> \; .
\ee
We shall also need the anomalous average
\be
\label{74}
\sgm_1(\br) \equiv \sgm_1(\br,\br) \; = \;
< \psi_1(\br) \psi_1(\br)> \; .
\ee
The absolute value $|\sgm_1(\br)|$ has the meaning of the density of
pair-correlated particles [43]. The total density
\be
\label{75}
\rho(\br) = \rho_0(\br) + \rho_1(\br)
\ee
is the average of the operator density (68). According to the normalization
conditions (20) and (21), the partial densities are normalized to the number
of corresponding particles,
\be
\label{76}
N_0 = \int \rho_0(\br)\; d\br \; , \qquad
N_1 = \int \rho_1(\br)\; d\br \; .
\ee
One more notation is the triple correlator
\be
\label{77}
\xi(\br,\br') \; \equiv \;
<\psi_1^\dgr(\br')\psi_1(\br')\psi_1(\br)> \; .
\ee
Using these definitions, the average of the correlation operator (59)
becomes
$$
<\hat X(\br,\br')> \; = \; \rho_1(\br')\eta(\br) +
\rho_1(\br,\br')\eta(\br') + \sgm_1(\br,\br') \eta^*(\br') +
\xi(\br,\br') \; .
$$

Finally, averaging Eq. (58) yields the evolution equation for the
condensate wave function
$$
i\; \frac{\prt}{\prt t}\; \eta(\br,t) = \left ( -\;
\frac{\nabla^2}{2m} + U - \mu_0 \right ) \eta(\br,t) 
$$
\be
\label{78}
+\; \int \Phi(\br-\br') \left [ \rho(\br')\eta(\br) +
\rho_1(\br,\br')\eta(\br') + \sgm_1(\br,\br')\eta^*(\br') +
\xi(\br,\br')\right ]\; d\br' \; .
\ee
This is the general equation for an arbitrary Bose-condensed system. This
equation can also be represented in a shorter, though not explicit, form
$$
i\; \frac{\prt}{\prt t}\; \eta(\br,t) = \left ( -\;
\frac{\nabla^2}{2m} + U - \mu_0 \right ) \eta(\br,t) +
\int \Phi(\br-\br')<\hat\psi^\dgr(\br')\hat\psi(\br')\hat\psi(\br)>
d\br' \; ,
$$
in which one has to substitute the shifted operator $\hat\psi=\eta+\psi_1$.

We may notice that Eq. (78) is invariant under the transformation
\be
\label{79}
\eta(\br,t) \; \longrightarrow \eta(\br,t)e^{i\al t} \; , \qquad
\psi_1(\br,t) \; \longrightarrow \psi_1(\br,t)e^{i\al t} \; , \qquad
\mu_0 \; \longrightarrow \mu_0 + \al \; .
\ee
This means that there exists a freedom in choosing the phase factor
$\exp(i\al t)$ of the condensate function. But the phase factor becomes
fixed by defining the Lagrange multiplier $\mu_0$. The arbitrariness in
the condensate phase implies that the stationary solutions for $\eta(\br,t)$
would be proportional to an undefined factor $e^{i\al t}$. By fixing the
multiplier $\mu_0$, we require that in equilibrium the condensate function
would not be dependent on time, that is,
\be
\label{80}
\frac{\prt}{\prt t}\; \eta(\br) = 0 \qquad ({\rm equilibrium}) \; .
\ee

In equilibrium, Eq. (78), according to condition (80), reduces to the
eigenvalue problem
$$
\mu_0\eta(\br) = \left [ -\; \frac{\nabla^2}{2m} + U(\br)\right ]
\eta(\br) 
$$
\be
\label{81}
+ \; \int \Phi(\br-\br') \left [ \rho(\br')\eta(\br) +
\rho_1(\br,\br')\eta(\br') + \sgm_1(\br,\br')\eta^*(\br') +
\xi(\br,\br') \right ] \; d\br' \; ,
\ee
defining, together with the normalization condition (76), the eigenfunction
$\eta(\br)$ and the eigenvalue $\mu_0$. Let us emphasize that without the
normalization condition (76) the condensate function cannot be uniquely
defined from Eq. (81). For example, if the system is uniform, with no
external potential $U\ra 0$, when $\eta(\br)=\eta$, $\rho(\br)=\rho$, and
$\eta=\sqrt{\rho_0}$, then Eq. (81) gives
\be
\label{82}
\mu_0 = \rho \Phi_0 + \int \Phi(\br) \left [ \rho_1(\br,0) +
\sgm_1(\br,0) + \frac{\xi(\br,0)}{\sqrt{\rho_0}} \right ] \; d\br \; ,
\ee
where $\Phi_0\equiv\int\Phi(\br)d\br$, or in the compact form
$$
\mu_0 = \frac{1}{\sqrt{\rho_0}} \; \int \Phi(\br)
<\hat\psi^\dgr(0)\hat\psi(0)\hat\psi(\br)>d\br \; .
$$
Expression (82) is valid for an arbitrary uniform equilibrium system with
an interaction potential $\Phi(\br)$. No approximations have been used in
obtaining Eq. (82).

In recent days, the physics of dilute Bose gases is intensively explored
(see the book [46] and review articles [31,47--49]). The interaction
potential for these gases is taken in the contact form
\be
\label{83}
\Phi(\br) = \Phi_0\dlt(\br) \; , \qquad
\Phi_0 \equiv 4\pi\; \frac{a_s}{m} \; ,
\ee
where $a_s$ is the scattering length. With this interaction potential,
the eigenvalue problem (81) for a nonuniform system becomes
$$
\mu_0 \eta(\br) = \left [ -\; \frac{\nabla^2}{2m} + U(\br) \right ]
\eta(\br) 
$$
\be
\label{84}
+ \; \Phi_0 \left\{ \left [ \rho(\br) + \rho_1(\br)\right ]\eta +
\sgm_1(\br)\eta^*(\br) + \xi(\br,\br)\right \} \; .
\ee
For a uniform system, $\mu_0$ is given by Eq. (82), which, in the case
of the contact potential (83), yields
\be
\label{85}
\mu_0 = \left ( \rho +\rho_1 +\sgm_1 +
\frac{\xi}{\sqrt{\rho_0}} \right ) \Phi_0 \; ,
\ee
where $\sgm_1\equiv\sgm_1(\br,\br)$ and $\xi\equiv\xi(\br,\br)$.

When one assumes that the Bose gas is dilute, being characterized by
the contact potential (83), the temperature is zero, and the interaction
is so weak that the condensate depletion can be neglected, so that all
particles are condensed, then $N_0=N$, $N_1=0$, and $\rho_1=\sgm_1=\xi=0$.
In this approximation, because of relation (44), the multiplier $\mu_0=\mu$
coincides with the system chemical potential. The eigenvalue problem (84)
reduces then to the Gross-Pitaevskii equation
$$
\left [ -\; \frac{\nabla^2}{2m} + U(\br) + \Phi_0|\eta(\br)|^2
\right ] \eta(\br) = \mu \eta(\br) \; .
$$
But when the condensate depletion is not negligible, then $\mu_0$,
according to Eq. (44), is not the same as $\mu$.

\section{Uniform Bose system}

Let us illustrate in more detail the application of the representative
ensemble with the grand Hamiltonian (52), to an equilibrium uniform
system. Then the field operators of uncondensed particles can be expanded
in plane waves,
\be
\label{86}
\psi_1(\br) =\sum_{k\neq 0} a_k \vp_k(\br) \; ,
\ee
where $\vp_k(\br)=e^{i\bk\cdot\br}/\sqrt{V}$. The condensate function
reduces to the constant $\eta=\sqrt{\rho_0}$, while the condensate
multiplier $\mu_0$ is given by Eq. (82).

In the momentum representation, the momentum distribution of particles
\be
\label{87}
n_k \equiv \; <a_k^\dgr a_k >
\ee
is usually termed the normal average, as compared to the anomalous average
\be
\label{88}
\sgm_k \equiv \; < a_k a_{-k} > \; .
\ee
The normal and anomalous density matrices (71) and (72) take the form of
the expansions
\be
\label{89}
\rho_1(\br,\br') =\sum_{k\neq 0} n_k \vp_k(\br) \vp_k^*(\br') \; ,
\qquad
\sgm_1(\br,\br') =\sum_{k\neq 0} \sgm_k \vp_k(\br) \vp_k^*(\br') \; ,
\ee
in which the properties
$$
<a_k^\dgr a_p> \; = \; \dlt_{kp} n_k \; , \qquad
<a_k a_p> \; = \; \dlt_{-kp}\sgm_k
$$
are taken into account. The diagonal elements of the matrices in Eq. (89)
give the densities
\be
\label{90}
\rho_1 = \rho_1(\br,\br) = \frac{1}{V} \; \sum_{k\neq 0} n_k \; ,
\qquad
\sgm_1 = \sgm_1(\br,\br) = \frac{1}{V} \; \sum_{k\neq 0} \sgm_k \; .
\ee
The interaction potential is assumed to allow the Fourier transformation
\be
\label{91}
\Phi(\br) = \frac{1}{V} \; \sum_k \Phi_k e^{i\bk\cdot\br} \; \qquad
\Phi_k = \int \Phi(\br) e^{-i\bk\cdot\br} \; d\br \; .
\ee
Then we find the following terms of the grand Hamiltonian (52). The
zero-order term (53) becomes
\be
\label{92}
H^{(0)} = \left ( \frac{1}{2}\; \rho_0 \Phi_0 - \mu_0 \right ) N_0 \; .
\ee
The first-order term $H^{1)}=0$ is automatically zero. The second-order
term (55) is
\be
\label{93}
H^{(2)} = \sum_{k\neq 0} \left [ \frac{k^2}{2m} +
\rho_0 (\Phi_0+\Phi_k) - \mu_1 \right ]  a_k^\dgr a_k +
\frac{1}{2} \; \sum_{k\neq 0} \rho_0 \Phi_k \left (
a_k^\dgr a_{-k}^\dgr + a_{-k} a_k \right ) \; .
\ee
For the third-order term (56), we have
\be
\label{94}
H^{(3)} = \sqrt{\frac{\rho_0}{V}} \; {\sum_{kp}}' \; \Phi_p
\left ( a_k^\dgr a_{k+p} a_{-p} + a_{-p}^\dgr a_{k+p}^\dgr a_k
\right ) \; ,
\ee
where the prime on the summation symbol implies that $\bk\neq 0$,
$\bp\neq 0$, $\bk+\bp\neq 0$. The fourth-order term (57) is
\be
\label{95}
H^{(4)} = \frac{1}{2V} \;
\sum_q {\sum_{kp}}' \; \Phi_q a_k^\dgr a_p^\dgr a_{p+q} a_{k-q} \; ,
\ee
where the prime shows that $\bk\neq 0$, $\bp\neq 0$, $\bp+{\bf q}\neq 0$,
$\bk-{\bf q}\neq 0$.

To be more specific, it is necessary to resort to some approximation. The
natural mean-field approximation for a system with broken gauge symmetry
is the HFB approximation. The latter is usually blamed to display an
unphysical gap in the spectrum, because of which it is qualified as gapful
(see detailed discussion in Refs. [7,31,32]). However, as is explained in
the Introduction, this defect comes into play only owing to the usage of
a nonrepresentative statistical ensemble. But for the representative
ensemble, with the grand Hamiltonian (52), there appear no such problems.
Below, we show this for the case of arbitrary temperature and interaction
potential $\Phi(\br)$.

We apply, in the standard way [40,45], the HFB approximation to the third-
and fourth-order products of the operators $a_k$. Then the third-order
term (94) becomes identically zero, because of the condition $<a_k>=0$,
\be
\label{96}
H^{(3)} = 0 \; .
\ee
And for the fourth-order term (95), we get
$$
H^{(4)} = \sum_{k\neq 0} \rho_1 \Phi_0 \left ( a_k^\dgr a_k \; - \;
\frac{1}{2}\; n_k \right ) 
$$
\be
\label{97}
+\; \frac{1}{V} \; \sum_{k,p\neq 0} \Phi_k \left [
n_{k+p} a_p^\dgr a_p + \frac{1}{2} \left (
\sgm_{k+p} a_p^\dgr a_{-p}^\dgr + \sgm_{k+p}^* a_{-p} a_p \right ) \;
- \; \frac{1}{2}\left ( n_{k+p} n_p + \sgm_{k+p} \sgm_p \right )
\right ] \; .
\ee

It is convenient to introduce the notations
\be
\label{98}
\om_k \equiv \frac{k^2}{2m} + \rho \Phi_0 + \rho_0 \Phi_k +
\frac{1}{V} \; \sum_{p\neq 0} n_p \Phi_{k+p} - \mu_1
\ee
and
\be
\label{99}
\Dlt_k \equiv \rho_0 \Phi_k + \frac{1}{V} \sum_{p\neq 0}
\sgm_p \Phi_{k+p} \; .
\ee
Since the interaction potential $\Phi(\br)=\Phi(-\br)$ is symmetric
and real, we have the properties $\sgm_k=\sgm_k^*=\sgm_{-k}$ and
$\Dlt_k=\Dlt_k^*=\Dlt_{-k}$.

Summing up all terms in the grand Hamiltonian (52), we obtain in the HFB
approximation
\be
\label{100}
H_{HFB} =  E_{HFB} + \sum_{k\neq 0} \om_k a_k^\dgr a_k +
\frac{1}{2} \sum_{k\neq 0} \Dlt_k \left ( a_k^\dgr a_{-k}^\dgr +
a_{-k} a_k \right ) \; ,
\ee
where the nonoperator term is
\be
\label{101}
E_{HFB} =  H^{(0)} \; - \; \frac{1}{2} \; \rho_1^2 \Phi_0 V \; - \;
\frac{1}{2V} \; \sum_{k,p\neq 0} \Phi_{k+p} \left ( n_k n_p +
\sgm_k \sgm_p \right ) \; .
\ee
The quadratic form (100) can be diagonalized by the Bogolubov canonical
transformation [3,4,45]
$$
a_k = u_k b_k + v_{-k}^* b_{-k}^\dgr \; .
$$
As a result, the grand Hamiltonian (100) is transformed to the Bogolubov
representation
\be
\label{102}
H_B = E_B + \sum_{k\neq 0} \ep_k b_k^\dgr b_k \; ,
\ee
in which
\be
\label{103}
E_B = E_{HFB}  + \frac{1}{2}\; \sum_{k\neq 0} (\ep_k - \om_k) \; .
\ee
For the operators $b_k$, one has the properties
\be
\label{104}
<b_k>\; = \; <b_k b_p> \; = \; 0 \; , \qquad
<b_k^\dgr b_p> \; = \; \dlt_{kp} \pi_k \; ,
\ee
with the momentum distribution of quasiparticles
\be
\label{105}
\pi_k \equiv \; < b_k^\dgr b_k > \; = \; \left ( e^{\bt\ep_k} -
1 \right )^{-1} = \frac{1}{2} \left [ {\rm coth}\left (
\frac{\ep_k}{2T}\right ) -1 \right ] \; .
\ee
The coefficient functions $u_k$ and $v_k$, and the spectrum $\ep_k$ are
defined by the Bogolubov - de Gennes equations
$$
(\om_k -\ep_k ) u_k + \Dlt_k v_k = 0 \; , \qquad
\Dlt_k u_k + (\om_k + \ep_k ) v_k = 0 \; ,
$$
with the normalization condition $u_k^2-v_k^2=1$. This gives
$$
u_k^2  = \frac{\om_k+\ep_k}{2\ep_k} \; , \qquad
v_k^2 = \frac{\om-\ep_k}{2\ep_k} \; .
$$
And the Bogolubov spectrum is
\be
\label{106}
\ep_k = \sqrt{\om_k^2 - \Dlt_k^2} \; .
\ee

The existence of the Bose-Einstein condensate, as is known, requires
that the spectrum (106) be gapless, such that
\be
\label{107}
\lim_{k\ra 0} \ep_k = 0 \; ,
\ee
with the stability condition $\ep_k\geq 0$. Then Eq. (106) yields
\be
\label{108}
\mu_1 = \rho \Phi_0 + \frac{1}{V} \;
\sum_{p\neq 0} ( n_p - \sgm_p) \Phi_p \; .
\ee
As will be shown in Sec. 10, this value is in exact agreement with the
Hugenholtz-Pines relation. This should be compared with the condensate
potential (82), which in the HFB approximation, when $\xi(\br,\br')=0$,
becomes
\be
\label{109}
\mu_0 = \rho \Phi_0 + \frac{1}{V} \;
\sum_{p\neq 0} ( n_p + \sgm_p) \Phi_p \; .
\ee
In the particular case of the contact potential (83), we get
\be
\label{110}
\mu_1 = (\rho + \rho_1 -\sgm_1) \Phi_0
\ee
and, respectively,
\be
\label{111}
\mu_0 = (\rho + \rho_1 + \sgm_1) \Phi_0 \; .
\ee
In any case, generally, $\mu_0\neq\mu_1$. The multipliers $\mu_0$
and $\mu_1$ become equal only in the limit of the asymptotically small
condensate depletion, which corresponds to the Bogolubov approximation
[1,2], when both $\rho_1$ and $\sgm_1$ in Eqs. (108) to (111) are
neglected. Then, clearly, $\mu_0=\mu_1=\rho_0\Phi_0$, and in view of
relation (44), $\mu_0=\mu_1=\mu$. However, as soon as the condensate
depletion is not negligible, $\mu_0\neq\mu_1$.

With the multiplier (108), Eq. (98) can be represented as
\be
\label{112}
\om_k = \frac{k^2}{2m} + \rho_0 \Phi_k + \frac{1}{V} \;
\sum_{p\neq 0} \left ( n_p \Phi_{k+p} - n_p \Phi_p +
\sgm_p \Phi_p \right ) \; .
\ee
For the long-wave limit of $\Dlt_k$, we introduce the notation
\be
\label{113}
\Dlt \equiv \lim_{k\ra 0} \Dlt_k \equiv mc^2 \; .
\ee
From Eq. (99) it follows
\be
\label{114}
\Dlt = \rho_0 \Phi_0 + \frac{1}{V} \;
\sum_{p\neq 0} \sgm_p \Phi_p \; .
\ee
Then the long-wave limit of spectrum (106) is explicitly of the phonon
type, $\ep\simeq ck$, as $k\ra 0$. According to the notation (113), the
sound velocity is
\be
\label{115}
c =\sqrt{\frac{\Dlt}{m} } \; .
\ee
Strictly speaking, the sound velocity is defined as $c=\sqrt{\Dlt/m^*}$, 
with an effective mass to be given in Sec. 10. For short-range 
interactions, $m^*\cong m$. Using Eqs. (113) and (114), we may rewrite 
Eq. (112) in the form
\be
\label{116}
\om_k = \frac{k^2}{2m} + mc^2 + \rho_0 (\Phi_k -\Phi_0) +
\int n_p (\Phi_{k+p} - \Phi_p) \; \frac{d\bk}{(2\pi)^3} \; .
\ee
For the particle momentum distribution (87), we find
\be
\label{117}
n_k = \frac{\om_k}{2\ep_k} \; {\rm coth}\left (
\frac{\ep_k}{2T} \right ) \; - \; \frac{1}{2} \; ,
\ee
while for the anomalous average (88), we obtain
\be
\label{118}
\sgm_k = -\; \frac{\Dlt_k}{2\ep_k} \; {\rm coth} \left (
\frac{\ep_k}{2T} \right ) \; .
\ee
Analyzing the behaviour of $n_k$ and $\sgm_k$ as functions of momentum
$k$, we find [15,32] that $|\sgm_k|\simeq n_k$ for $k\ra 0$, while
$|\sgm_k|\gg n_k$ for large $k$. Therefore in no sense the anomalous
average $\sgm_k$ can be neglected, as compared to the normal average
$n_k$.

The derived equations simplify for the contact potential (83). Then
$\Phi_k=\Phi_0$, $\Dlt_k=\Dlt$,
$$
\Dlt=(\rho_0 + \sgm_1) \Phi_0 \; ,
$$
and spectrum (106) takes the classical Bogolubov form
\be
\label{119}
\ep_k = \sqrt{(ck)^2 + \left ( \frac{k^2}{2m}\right )^2 } \; .
\ee

The grand potential (36) in the HFB approximation is
\be
\label{120}
\Om = E_B + TV \int \ln \left ( 1  - e^{-\bt\ep_k}
\right ) \; \frac{d\bk}{(2\pi)^3 } \; .
\ee
From this and other equations, obtained above, all thermodynamic
characteristics can be calculated.

\section{Condensate and Superfluid Fractions}

We shall concentrate our attention on the most interesting characteristics
of the Bose-condensed system, on the condensate and superfluid fractions.
The condensate fraction
\be
\label{121}
n_0 =  1 \; - \; \frac{\rho_1}{\rho}
\ee
can be found by calculating the density $\rho_1$ of uncondensed particles,
given by the integral
\be
\label{122}
\rho_1 = \int n_k \; \frac{d\bk}{(2\pi)^3} \; .
\ee

To define the superfluid fraction, one considers the reaction of the
system to the boost with a velocity $\bv$. In the laboratory frame,
the field operators of a moving system are represented by means of the
Galilean transformation
\be
\label{123}
\hat\psi_v(\br,t) = \hat\psi(\br-\bv t,t) \exp\left \{ i \left (
m\bv \cdot\br \; - \; \frac{mv^2}{2}\; t\right )\right \} \; .
\ee
The Hamiltonian $H_v=H[\hat\psi_v]$, in terms of the new field operators
(123), is expressed through the Hamiltonian $H=H[\hat\psi]$, in terms of
the old operators $\hat\psi$, as
\be
\label{124}
H_v = H + \int \hat\psi^\dgr(\br) \left ( - i\bv\cdot\nabla +
\frac{mv^2}{2} \right ) \hat\psi(\br)\; d\br \; .
\ee
The total system momentum operator becomes
\be
\label{125}
\hat{\bf P}_v = \hat{\bf P} + Nm\bv \; ,
\ee
where
\be
\label{126}
\hat{\bf P}  = \int \hat\psi^\dgr(\br) (-i\nabla)\hat\psi(\br)\; d\br
\ee
is the old momentum operator. The average total momentum of the moving
system is
\be
\label{127}
<\hat{\bf P}_v>_v \; \equiv \;
{\rm Tr}\hat\rho_v \hat{\bf P}_v \; ,
\ee
with the statistical operator
\be
\label{128}
\hat\rho_v \equiv \frac{e^{-\bt H_v}}{{\rm Tr}e^{-\bt H_v}} \; .
\ee

The superfluid fraction is defined as
\be
\label{129}
n_s \equiv \frac{1}{3mN}\; \lim_{v\ra 0} \; \frac{\prt}{\prt\bv}\;
\cdot <\hat{\bf P}_v>_v \; .
\ee
It is possible to show (see, e.g., the detailed derivation in Refs.
[47,49]) that the superfluid fraction (129) for an arbitrary system can be
represented in the form
\be
\label{130}
n_s =  1 \; - \; \frac{2Q}{3T} \; ,
\ee
in which
\be
\label{131}
Q \equiv \frac{\Dlt^2(\hat{\bf P})}{2mN}
\ee
is the {\it dispersed heat}, with the momentum dispersion
$$
\Dlt^2(\hat{\bf P}) \equiv \; <\hat{\bf P}^2> - < \hat{\bf P}>^2 \; ,
$$
where the averages are with respect to the grand Hamiltonian $H[\hat\psi]$.

Formula (130) for the superfluid fraction is valid for any system,
equilibrium or nonequilibrium, uniform or not. For a system in equilibrium,
the total momentum is zero, $<\hat{\bf P}>=0$, so that the dispersed heat
(131) is
\be
\label{132}
Q = \frac{<\hat{\bf P}^2>}{2mN} \; .
\ee
For a uniform system,
\be
\label{133}
<\hat{\bf P}^2> \; = \; \sum_{kp} \left (\bk \cdot {\bf p} \right )
<\hat n_k \hat n_p> \; ,
\ee
where $\hat n_k\equiv a_k^\dgr a_k$. In the HFB approximation,
\be
\label{134}
<\hat n_k \hat n_p> \; = \; n_k n_p + \dlt_{kp} n_k (1 + n_k) +
\dlt_{-kp} \sgm_k^2 \; .
\ee
Then the dispersed heat (132) becomes
\be
\label{135}
Q = \frac{1}{\rho} \; \int \frac{k^2}{2m} \left (
n_k + n_k^2 - \sgm_k^2 \right ) \frac{d\bk}{(2\pi)^3} \; .
\ee
Substituting here expressions (117) and (118), and using the equality
${\rm cosh}x-1=2{\rm sinh}^2(x/2)$, so that
$$
n_k + n_k ^2 - \sgm_k^2 = \frac{1}{4{\rm sinh}^2(\bt\ep_k/2)} \; ,
$$
we get from Eq. (135)
\be
\label{136}
Q = \frac{1}{(4\pi)^2m\rho} \; \int_0^\infty \;
\frac{k^4\; dk}{{\sinh}^2(\bt\ep_k/2)} \; .
\ee
Note again the importance of the anomalous average $\sgm_k$. If one would
omit the latter in the dispersed heat (135), one would get a senseless
divergent quantity, while accurately taking account of $\sgm_k$ results in
the well-defined convergent integral (136).

To demonstrate explicitly the behaviour of the condensate and superfluid
fractions, let us resort to the contact potential (83). Then Eq. (99) yields
\be
\label{137}
\Dlt_k =\Dlt = mc^2 \; ,
\ee
and Eq. (116) gives
\be
\label{138}
\om_k = \frac{k^2}{2m} + mc^2 \; .
\ee
Equation (115), defining the sound velocity, can be written as
\be
\label{139}
mc^2 = (\rho_0 +\sgm_1) \Phi_0 \; ,
\ee
where
\be
\label{140}
\sgm_1 = \int \sgm_k \; \frac{d\bk}{(2\pi)^3} \; .
\ee

The density of uncondensed particles (122) can be represented in the form
\be
\label{141}
\rho_1 = \frac{(mc)^3}{3\pi^2} \left \{ 1 + \frac{3}{2\sqrt{2}} \;
\int_0^\infty \; \left (\sqrt{1+x^2}-1\right )^{1/2} \left [
{\rm coth}\left ( \frac{mc^2}{2T}\; x\right ) -1 \right ]\; dx
\right \} \; ,
\ee
which is a well-defined convergent integral.

The anomalous average (140) can be written as a sum of two parts
\be
\label{142}
\sgm_1 = \sgm_0  - \int \; \frac{mc^2}{2\ep_k} \left [
{\rm coth}\left ( \frac{\ep_k}{2T} \right ) -1 \right ] \;
\frac{d\bk}{(2\pi)^3} \; ,
\ee
in which
\be
\label{143}
\sgm_0 \equiv - \Dlt \int \; \frac{1}{2\ep_k} \;
\frac{d\bk}{(2\pi)^3} \; .
\ee
The second integral in Eq. (142) is convergent. But the integral in Eq. (143)
ultravioletly diverges. However, this divergence is known to be unphysical,
being simply caused by the usage of the contact interaction potential. This
and other similar divergences could be removed by employing more realistic
interaction potentials. For example, a common choice is a Gaussian type
potential [50]. Another known way of removing such divergences is by using
the analytic regularization in one of its variants, such as the subtraction
scheme, zeta regularization, or dimensional regularization (see details in
Ref. [31]). This kind of regularization is asymptotically exact for weak
interactions and is universal in the sense that it applies to any short-range
potential with a scattering length $a_s$ [31]. Using the dimensional
regularization in its region of validity, we obtain
$$
\int \frac{1}{2\ep_k} \; \frac{d\bk}{(2\pi)^3} =  - \;
\frac{m}{\pi^2}\; \sqrt{m\rho_0\Phi_0} \; .
$$
Then Eq. (143) results in
\be
\label{144}
\sgm_0 = \frac{(mc)^2}{\pi^2} \; \sqrt{m\rho_0\Phi_0} \; .
\ee
The anomalous average (142) can be represented as
\be
\label{145}
\sgm_1 = \sgm_0 \; - \; \frac{(mc)^3}{2\sqrt{2}\;\pi^2} \;
\int_0^\infty \; \frac{(\sqrt{1+x^2}-1)^{1/2}}{\sqrt{1+x^2}} \left [
{\rm coth}\left (\frac{mc^2}{2T}\; x \right ) -1 \right ] \; dx \; .
\ee
And for the dispersed heat (136), we obtain
\be
\label{146}
Q = \frac{(mc)^5}{\sqrt{2}\; (2\pi)^2 m\rho} \; \int_0^\infty \;
\frac{(\sqrt{1+x^2}-1)^{3/2}\; xdx}
{\sqrt{1+x^2}\; {\rm sinh}^2(mc^2x/2T)} \; .
\ee

At low temperatures, such that
$$
\frac{T}{mc^2} \ll 1 \; ,
$$
equations (141), (145), and (146) yield
$$
\rho_1 \simeq \frac{(mc)^3}{3\pi^2} + \frac{(mc)^3}{12} \; \left (
\frac{T}{mc^2}\right )^2 \; , \qquad
\sgm_1 \simeq \sgm_0 \; - \; \frac{(mc)^3}{12}\left (
\frac{T}{mc^2}\right )^2 \; ,
$$
$$
Q \simeq \frac{\pi^2(mc)^5}{15m\rho} \; \left (
\frac{T}{mc^2}\right )^5 \; .
$$
Therefore, the condensate fraction (121) behaves as
\be
\label{147}
n_0 \simeq 1 \; - \; \frac{(mc)^3}{3\pi^2\rho} \; - \;
\frac{(mc)^3}{12\rho}\; \left ( \frac{T}{mc^2} \right )^2 \; .
\ee
And the superfluid fraction (130) becomes
\be
\label{148}
n_s \simeq 1 \; - \; \frac{2\pi^2(mc)^3}{45\rho} \;
\left ( \frac{T}{mc^2}\right )^4 \; .
\ee
When the interaction is weak, or the condensate fraction $n_0$ tends to
zero at the critical temperature $T_c$, as a results of which, $c\ra 0$,
so that
$$
\frac{mc^2}{T_c} \ll 1 \; ,
$$
then Eqs. (141), (145), and (146) lead to
$$
\rho_1 \simeq \rho \left ( \frac{T}{T_c} \right )^{3/2} +
\frac{(mc)^3}{3\pi^2} \; , \qquad \sgm_1 \simeq \sgm_0 \; - \;
\frac{m^2cT}{2\pi} \; ,
$$
$$
Q \simeq \frac{3}{2}\; T \; \left [ \left ( \frac{T}{T_c}
\right )^{3/2} - \; \frac{\zeta(1/2)}{\zeta(3/2)} \left (
\frac{T}{T_c} \right )^{1/2}
\frac{mc^2}{T_c} \right ] \; ,
$$
with the critical temperature
$$
T_c = \frac{2\pi}{m} \left [ \frac{\rho}{\zeta(3/2)}\right ]^{2/3}
$$
coinciding with that for the ideal Bose gas, as it should be in the case
of the mean-field approximation with the contact potential (83). Here
$\zeta(\cdot)$ is the Riemann zeta function. Then the condensate fraction
tends to zero at $T_c$ as
\be
\label{149}
n_0 \simeq 1 \; - \; \left ( \frac{T}{T_c} \right )^{3/2} - \;
\frac{(mc)^3}{3\pi^2\rho} \; ,
\ee
together with the superfluid fraction
\be
\label{150}
n_s \simeq 1 \; - \; \left ( \frac{T}{T_c}\right )^{3/2} + \;
\frac{\zeta(1/2)}{\zeta(3/2)}\; \left ( \frac{T}{T_c}\right )^{1/2}
\frac{mc^2}{T_c} \; .
\ee
Both, $n_0$ and $n_s$, tend to zero from the left, demonstrating the
second-order phase transition at $T_c$. The point why in the present case
$T_c$ has to be the same as the ideal Bose gas critical temperature is
explained in the Appendix C.

\section{Green Function Equations}

In order to conclude the proof of the complete self-consistency of the
developed approach, based on the introduced representative ensemble, we need
to show that for an arbitrary Bose system there exist the Green-function
equations of the usual form and that the Hugenholtz-Pines relation follows
for any uniform system.

The first-order Green function is defined in the standard way [3,4,7,9] as the
matrix $G(12)=[G_{\al\bt}(12)]$ with the elements
$$
G_{11}(12) = - i<\hat T\psi_1(1)\psi_1^\dgr(2)> \; , \qquad
G_{12}(12) = - i<\hat T\psi_1(1)\psi_1(2)> \; ,
$$
\be
\label{151}
G_{21}(12) = - i<\hat T\psi_1^\dgr(1)\psi_1^\dgr(2)> \; , \qquad
G_{22}(12) = - i<\hat T\psi_1^\dgr(1)\psi_1(2)> \; ,
\ee
in which the common abbreviation is employed, denoting the set $\{\br_j,t_j\}$
by a single number $j$, and $\hat T$ is the chronological operator.

For what follows, we shall need the notation for the triple field operator
$$
\Psi(123) \equiv \psi_1(1)\psi_1(2)\psi_1^\dgr(3) +
\eta(1)\psi_1(2)\psi_1^\dgr(3) + \psi_1(1)\eta(2)\psi_1^\dgr(3)
$$
\be
\label{152}
+\; \psi_1(1)\psi_1(2)\eta^*(3) +
\eta(1)\eta(2)\psi_1^\dgr(3) + \eta(1)\psi_1(2)\eta^*(3) +
\psi_1(1)\eta(2)\eta^*(3) \; .
\ee
One may notice that
$$
\Psi(123) = \hat\psi(1) \hat\psi(2)\hat\psi^\dgr(3)  -
\eta(1)\eta(2)\eta^*(3) \; ,
$$
where $\hat\psi=\eta+\psi_1$ is the shifted field operator. However,
for practical application, we need the explicit form (152).

The second-order Green function is a matrix
$B(1234)=[B_{\al\bt}(1234)]$ with the elements
$$
B_{11}(1234) = -<\hat T\Psi(123)\psi_1^\dgr(4)> \; , \qquad
B_{12}(1234) = -<\hat T\Psi(123)\psi_1(4)> \; ,
$$
\be
\label{153}
B_{21}(1234) = -<\hat T\Psi^\dgr(123)\psi_1^\dgr(4)> \; , \qquad
B_{22}(1234) = -<\hat T\Psi^\dgr(123)\psi_1(4)> \; .
\ee

For the general form of a retarded interaction potential
\be
\label{154}
\Phi(12) \equiv \Phi(\br_1 - \br_2) \dlt(t_{12}+0) \; ,
\ee
in which $t_{12}=t_1-t_2$, the self-energy is introduced through the equation
\be
\label{155}
\int \Sigma(13) G(32)\; d(3) = i \int \Phi(13) B(1332)\; d(3) \; .
\ee
The equation for $G(12)$, by defining the inverse propagator
\be
\label{156}
G^{-1}(12) \equiv \left ( \hat\tau \; i \;
\frac{\prt}{\prt t_1} + \frac{\nabla_1^2}{2m} \; - \; U(1) +
\mu_1 \right ) \dlt(12) \; \hat 1 - \Sgm(12) \; ,
\ee
in which
\begin{eqnarray}
\nonumber
\hat\tau \equiv \left [ \begin{array}{rr}
1 & 0 \\
0 & -1 \end{array}  \right ]  \; , \qquad
\hat 1 \equiv \left [ \begin{array}{rr}
1 & 0 \\
0 & 1  \end{array} \right ] \; ,
\end{eqnarray}
can be represented in the form
\be
\label{157}
\int G^{-1}(13) G(32) \; d(3) = \dlt(12)\; \hat 1 \; .
\ee

To solve Eq. (157), one may invoke perturbation theory, starting with
an available approximate Green function $G_{app}$, corresponding to an
approximate $\Sigma_{app}$, such that
\be
\label{158}
\int G_{app}^{-1}(13) G_{app}(32)\; d(3) = \dlt(12)\hat 1 \; .
\ee
Then Eqs. (157) and (158) can be transformed to the Dyson representation
\be
\label{159}
G(12) = G_{app}(12) + \int G_{app}(13) \left [
\Sigma(34) - \Sigma_{app}(34)\right ] G(42)\; d(34) \; ,
\ee
which is convenient for using perturbation theory.

The above equations for the Green functions are valid for an arbitrary
nonequilibrium and nonuniform Bose system. If the considered system is
equilibrium and uniform, one passes to the Fourier transforms $G(\bk,\om)$
and $\Sigma(\bk,\om)$, employing the symmetry properties
$$
G_{11}(\bk,-\om) = G_{22}(\bk,\om) \; , \quad
G_{12}(\bk,-\om) = G_{21}(\bk,\om) = G_{12}(\bk,\om) \; ,
$$
\be
\label{160}
G_{\al\bt}(-\bk,\om) = G_{\al\bt}(\bk,\om)
\ee
and, respectively,
$$
\Sigma_{11}(\bk,-\om) = \Sigma_{22}(\bk,\om) \; , \qquad
\Sigma_{12}(\bk,-\om) =\Sigma_{21}(\bk,\om) =\Sigma_{12}(\bk,\om) \; ,
$$
\be
\label{161}
\Sigma_{\al\bt}(-\bk,\om) =\Sigma_{\al\bt}(\bk,\om) \; .
\ee
A detailed discussion of these relations can be found in Ref. [4].

Let us introduce the notation
\be
\label{162}
D(\bk,\om) \equiv \Sigma_{12}^2 (\bk,\om) -
G_{11}^{-1}(\bk,\om) G_{11}^{-1}(\bk,-\om) \; ,
\ee
in which
$$
G_{11}^{-1}(\bk,\om) = \om \; - \; \frac{k^2}{2m} \; - \;
\Sigma_{11}(\bk,\om) + \mu_1 \; .
$$
Then the solution to Eq. (157) can be written as
\be
\label{163}
G_{11}(\bk,\om) =
\frac{\om+k^2/2m+\Sigma_{11}(\bk,\om)-\mu_1}{D(\bk,\om)} \; , \qquad
G_{12}(\bk,\om) = -\; \frac{\Sigma_{12}(\bk,\om)}{D(\bk,\om)} \; ,
\ee
with the denominator (162).

For the Green functions $G_{\al\bt}(\bk,0)$ at zero energy there is the
Bogolubov theorem [4] rigorously proving the validity of the inequalities
\be
\label{164}
| G_{11}(\bk,0)| \geq \frac{mn_0}{2k^2} \; ,
\ee
\be
\label{165}
| G_{11}(\bk,0) - G_{12}(\bk,0) | \geq \frac{mn_0}{k^2} \; .
\ee
From Eq. (163), we have
$$
G_{11}(\bk,0) =
\frac{k^2/2m+\Sigma_{11}(\bk,0)-\mu_1}{D(\bk,0)} \; , \qquad
G_{12}(\bk,0) = -\; \frac{\Sigma_{12}(\bk,0)}{D(\bk,0)} \; ,
$$
where
$$
D(\bk,0) = \Sigma_{12}^2(\bk,0) - \left [ \frac{k^2}{2m} +
\Sigma_{11}(\bk,0) - \mu_1 \right ]^2 \; .
$$
Consequently,
$$
G_{11}(\bk,0) - G_{12}(\bk,0) = \left [ \mu_1 \; - \; \frac{k^2}{2m} +
\Sigma_{12}(\bk,0) - \Sigma_{11}(\bk,0) \right ]^{-1} \; .
$$
Using this in Eq. (165), we get
$$
\left | \mu_1 \; - \; \frac{k^2}{2m} + \Sigma_{12}(\bk,0) -
\Sigma_{11}(\bk,0) \right | \leq \frac{k^2}{mn_0} \; .
$$
Setting in the latter equation $\bk\ra 0$, we come to the Hugenholtz-Pines
relation
\be
\label{166}
\mu_1 = \Sigma_{11}(0,0) - \Sigma_{12}(0,0) \; .
\ee

Note that in the HFB approximation we have
$$
\Sigma_{11}(0,0) = (\rho+\rho_0)\Phi_0 + \frac{1}{V}\;
\sum_{p\neq 0} n_p \Phi_p \; , \qquad
\Sigma_{12}(0,0) = \rho_0 \Phi_0 + \frac{1}{V} \;
\sum_{p\neq 0} \sgm_p\Phi_p \; .
$$
Therefore Eq. (166) gives exactly the form of $\mu_1$ in Eq. (108).

Relation (166) guarantees that the system spectrum is gapless and of the
phonon character. The spectrum $\ep_k$ is given by the zeros of the Green
functions (163), that is, by the equation
\be
\label{167}
D(\bk,\ep_k) = 0 \; ,
\ee
with Eq. (162). Equation (167) can be rewritten as
\be
\label{168}
\ep_k = \frac{1}{2} \; \left [ \Sigma_{11}(\bk,\ep_k) -
\Sigma_{22}(\bk,\ep_k) \right ] +
\sqrt{\om_k^2- \Sigma_{12}^2(\bk,\ep_k)} \; ,
\ee
where the notation
$$
\om_k \equiv \frac{k^2}{2m} + \frac{1}{2} \; \left [
\Sigma_{11}(\bk,\ep_k) + \Sigma_{22}(\bk,\ep_k) \right ] - \mu_1
$$
is used. From Eq. (168), it follows that $\ep_k\ra 0$, as $k\ra 0$. Moreover,
when the system is isotropic, its self-energy $\Sigma_{\al\bt}(\bk,\ep_k)$
depends only on the scalar $k^2$. This implies the asymptotic, as $k\ra 0$,
expansion
$$
\Sigma_{\al\bt}(\bk,\ep_k) \simeq \Sigma_{\al\bt}(0,0) +
{\Sigma}'_{\al\bt}\;  k^2 \; ,
$$
in which
$$
{\Sigma'}_{\al\bt} \equiv \lim_{k\ra 0} \; \frac{\prt}{\prt k^2} \;
\Sigma_{\al\bt}(\bk,\ep_k) \; .
$$
Using this expansion in spectrum (168) gives
\be
\label{169}
\ep_k \simeq ck \; , \qquad
c \equiv \sqrt{\frac{1}{m^*}\; \Sigma_{12}(0,0)} \; ,
\ee
which is the phonon spectrum, with the sound velocity $c$, where the
effective mass is
\be
\label{170}
m^* \equiv \frac{m}{1+m(\Sigma'_{11}+\Sigma'_{22}-2\Sigma'_{12})} \; .
\ee

It is useful to compare the Lagrange multipliers $\mu_0$ and $\mu_1$
given by their general expressions (82) and (166), which are valid for
an arbitrary equilibrium uniform Bose system. These expressions are exact,
with no approximations being involved. As is seen, anomalous averages enter
$\mu_0$ with the sign plus, while the anomalous self-energy enters $\mu_1$
with the sign minus. This is why, in general, $\mu_0$ cannot coincide with
$\mu_1$. The HFB forms (108) and (109), or (110) and (111), are particular
illustrations. It is easy to check that $\mu_0$ coincides with $\mu_1$
solely in the Bogolubov limit of asymptotically weak interactions, when
$\Sigma_{11}(0,0)\ra(\rho+\rho_0)\Phi_0$, $\Sigma_{12}(0,0)\ra\rho_0\Phi_0$,
and $\mu_0\ra\mu_1\ra\mu=\rho\Phi_0$. But in any higher-order approximation,
$\mu_0\neq\mu_1$. The assumption that $\mu_0$ would be the same as $\mu_1$
would make the theory not self-consistent and would return us back to the
Hohenberg-Martin dilemma of conserving versus gapless theories.

\section{Discussion}

The main message of this paper is the necessity of employing representative
ensembles for correctly describing statistical systems. A representative
ensemble takes into account all imposed constraints and additional conditions
that allow for a unique description of the considered system. It is only using
a representative ensemble makes the theory self-consistent.

In the Bose system with broken global gauge symmetry, realized by the
Bogolubov shift, there are two particle components, corresponding to condensed
and uncondensed particles, with two related normalization conditions for $N_0$
and $N_1$. This requires to introduce two Lagrange multipliers, $\mu_0$ and
$\mu_1$, which makes the theory completely self-consistent in any approximation.

It is worth recalling that the introduction of several Lagrange multipliers
is rather common for spin systems. There, the order parameter is the average
spin, which, generally, is a three-component vector. The role of the effective
chemical potential for spin systems is played, as is well known, by an external
magnetic field, which is also a three-component vector. Hence, the number of
effective chemical potentials for spin systems is equal to the number of
components in the order parameter. Only then one is able to unambiguously
define the average spin.

The suggested approach, introducing two Lagrange multipliers, does not
contradict our physical understanding that the standard experiments fix, as
independent variables, temperature, volume, and the total number of particles.
To emphasize this once again, let us turn to the definition of the grand
thermodynamic potential (36), from which it follows that it is a function
$\Om = \Om(T,V,\mu_0,\mu_1)$, so that the free energy (42) is a function
$F=F(T,V,N_0,N_1)$. The Lagrange multiplier $\mu_0$ is defined from the 
stability condition, yielding $\mu_0 = \mu_0(T,V,N_0,N_1)$, according to 
Eqs. (81), (82), (84), (85), (109), and (111). The Lagrange multipler 
$\mu_1$ satisfies the Hugenholtz-Pines theorem, which gives
$\mu_1=\mu_1(T,V,N_0,N_1)$, in agreement with Eqs. (108), (110), and (166). 
The number of uncondensed particles is found from the direct calculation of 
the average (21) expressed through Eqs. (90), (122), (141), and like that, 
resulting in $N_1= N_1(T,V,N_0,N)$. From here, since $N_0=N-N_1$, it follows 
that $N_0 = N_0(T,V,N)$. Substituting it back to $N_1$, one has
$N_1=N_1(T,V,N)$. Using these in the expressions for $\mu_0$ and $\mu_1$, 
one gets $\mu_0=\mu_0(T,V,N)$ and $\mu_1=\mu_1(T,V,N)$. Then, from Eq. (44), 
it is evident that $\mu=\mu(T,V,N)$, or, inverting the latter relation, one 
has $N=N(T,V,\mu)$. Using this, the Lagrange multipliers $\mu_0$ and $\mu_1$ 
can be expressed as functions $\mu_0=\mu_0(T,V,\mu)$ and $\mu_1=\mu_1(T,V,\mu)$.
Substituting this into the grand potential, we have $\Om = \Om(T,V,\mu)$,
in line with Eq. (48). Respectively, the free energy becomes a function
$F=F(T,V,N)$, in accordance with Eq. (45).

Thus, at the end, we work with the standard variables $T,V$, and $N$,
which are usually fixed in experiments. All observable quantities are
also expressed through these variables. So that the suggested approach
is absolutely self-consistent, mathematically correct, and in agreement
with physics.

Several words are to be said with regard to the phase-transition order 
of Bose-Einstein condensation. This transition is known to be of {\it
second-order}, which is firmly based on several facts. First, there exists
a general explanation, independent of the coupling strength, that this
transition is of second-order. This can be found in the book by Patashinsky
and Pokrovsky [51] (Chapter X, Section 2). As is also well known, the
Hamiltonian of Bose systems is mathematically equivalent to what in quantum
theory is termed the $\vp^4$ model. The phase transition in this model has
been studied in numerous works using the renormalization group approach,
exhibiting the second-order transition [52]. The superfluid transition in
liquid $^4$He, which is believed to be accompanied by the Bose-Einstein
condensation, is also a second-order transition. A large body of experimental
data on measuring the continuous temperature dependence of the condensate
fraction in superfluid helium has been summarized by Wirth and Hallock [53].
There exists abundant literature, both theoretical and experimental, on
Bose-Einstein condensation in dilute trapped gases (see review works
[46--49]) and references therein) and there are several computer simulations
of this process [54--56]. Though Bose condensation in traps is smeared out
by finite-size effects, the subsequent increase of the number of particles
unambiguously demonstrates that the condensation approaches the standard
second-order phase transition. In addition, there have been many Monte Carlo
calculations for uniform Bose systems with various interaction potentials.
All these calculations, summarized in the review articles [57--59], clearly
prove that Bose-Einstein condensation is a second-order transition. So, the
second order of the Bose condensate transition has been established without
any doubt. This especially concerns the general theoretical investigations
[51,52] and rigorous Monte Carlo calculations [57,58].

The main idea of the present paper is the necessity of using representative
ensembles for describing Bose-condensed systems. This implies that proper
allowance must be made for all conditions which uniquely define the employed
field variables. Here the consideration has been based on the classical
Bogolubov approach [1--4] introducing two field variables, the condensate
function $\eta$ and the operator of uncondensed particles $\psi_1$. By their
definition, these variables are {\it independent} of each other. For a uniform
system, the condensate corresponds to the zero-momentum state, while this
state is excluded from the description of uncondensed particles. This is
evident from definition (86) of $\psi_1$, where $\bk\neq 0$. The variables
$\eta$ and $\psi_1$ are also {\it orthogonal} to each other, in agreement with
definition (17), which becomes obvious from Eq. (86), since
\be
\label{171}
\frac{1}{\sqrt{V}}\; \int \psi_1(\br)\; d\br = \sum_{k\neq 0}
a_k \dlt_{k0} = 0 \; .
\ee
For so introduced independent orthogonal variables, it is necessary to
define two normalization conditions and, respectively, two Lagrange
multipliers controlling these normalization conditions. Only then there can
be the assurance that the theory will be self-consistent in any calculations.

Of course, if the field variables are introduced in a different way, with
some other conditions, this would require to define another ensemble, with
the appropriate Lagrange multipliers, whose number could also be different.
For example, Hugenholtz and Pines [9], as well as later Gavoret and Nozieres
[60], when deriving the Hugenholtz-Pines relation on the basis of
thermodynamic properties, {\it did not use} the standard grand ensemble.
Any attentive reader can immediately infer from the original works [9,60]
that these authors have used a different ensemble. They treat a uniform
equilibrium system, defining the number of condensed particles $N_0$ by
minimizing the {\it internal energy} at zero temperature, which corresponds
to the minimization of {\it free energy} at finite temperatures. After
defining in that way $N_0=N_0(\rho,T)$, the latter is explicitly substituted
into the effective Hamiltonian $\hat H-\mu\hat N_1$, where $\hat N_1$ is the
operator of {\it uncondensed} particles, but not of the total number of
particles, as it would be in the standard grand Hamiltonian. Since
$N_0=N_0(\rho,T)$ has been explicitly substituted everywhere, one needs
the sole Lagrange multiplier for the normalization of {\it uncondensed}
particles. But mathematically this is absolutely equivalent to the
introduction of an additional Lagrange multiplier $\mu_0$, as is done
in the present paper, {\it before} substituting $N_0(\rho,T)$, which is
defined later from the corresponding normalization condition. These ways,
as is absolutely clear, are equivalent, but the latter method is more
convenient for more general cases of nonuniform or nonequilibrium systems.

Another example is given by the approach advanced by Faddeev and Popov [61]
and used later by Popov [12--14]. They define the field operator
\be
\label{172}
\psi(\br) =\sqrt{\rho_0} + \psi_{FP}(\br) \; ,
\ee
in which the second part in the right-hand side,
\be
\label{173}
\psi_{FP}(\br) = \frac{1}{\sqrt{V}}\; \sum_k a_k e^{i\bk\cdot\br} \; ,
\ee
includes the term $a_0\neq 0$ with $\bk=0$. This is contrary to the Bogolubov
field operator of uncondensed particles (86), not containing this
zero-momentum term. The Fourier transform $c_k$ of the operator
$$
\psi(\br) = \frac{1}{\sqrt{V}} \; \sum_k c_k e^{i\bk\cdot\br}
$$
is connected with the Fourier transform of $\psi_{FP}$ by the equation
$$
c_k = \dlt_{0k}\; \sqrt{N_0} + a_k \; .
$$
For the zero-momentum state, one has
$$
c_0 =\sqrt{N_0} + a_0 \; .
$$
Recall that in the Bogolubov case, one would have $c_0=\sqrt{N_0}$. Thus,
in the Faddeev-Popov approach, the condensate is not completely separated from
uncondensed particles, but $\psi_{FP}$ does contain the zero-momentum term
$a_0$. In other words, $\psi_{FP}$ is not independent from $\rho_0$. Moreover,
the Faddeev-Popov representation (172) for $\psi$ consists of two parts that
are not orthogonal to each other,
$$
\frac{1}{\sqrt{V}} \; \int \psi_{FP}(\br)\; d\br = a_0 \neq 0 \; ,
$$
which is contrary to the Bogolubov case (171). Hence, $\psi_{FP}$ is neither
independent of $\sqrt{\rho_0}$ no orthogonal to it, but both of them define
the sole variable (172). As far as, in the Faddeev-Popov approach, $a_0\neq
0$, the latter yields the interference terms in physical operators. For
instance, here the number-of-particle operator becomes
\be
\label{174}
\hat N \equiv \int \psi^\dgr(\br)\psi(\br)\; d\br = N_0 +
\sum_k a_k^\dgr a_k +\sqrt{N_0}\left ( a_0^\dgr + a_0\right ) \; .
\ee
This operator is normalized to the total number of particles, so that
\be
\label{175}
N \;  =\; <\hat N> \; = \; N_0 + \sum_k <a_k^\dgr a_k> \; ,
\ee
which requires $<a_0>=0$. Faddeev and Popov in their original paper [61]
emphasized that their representation (172) is principally different from the
Bogolubov shift (16) and discussed in detail the corresponding differences.
Because in the Faddeev-Popov approach there is a single independent variable
$\psi$, with the normalization condition (175), the appropriate representative
ensemble here is the standard grand ensemble, with the grand Hamiltonian $\hat
H-\mu\hat N$, though the number-of-particle operator (174) here is different
from the Bogolubov form (24). More complicated forms of physical operators and
the necessity to comply with relation (172) at each step of any calculational
procedure make the usage of the Faddeev-Popov approach more complicated and
quite inconvenient for mean-field type approximations. However, needless to say
that a theory with one independent variable and, respectively, with one
Lagrange multiplier, is mathematically equivalent to the theory with two
independent variables and two Lagrange multipliers. Which representation to
choose is rather a matter of convenience.

It is even admissible to introduce no Lagrange multipliers and to deal with
the canonical Gibbs ensemble. But the problem with the latter is that then one
has to keep the total number of particles fixed not merely on the average but
exactly at each step of any calculational procedure. This requires to work in
a restricted Fock space, where the number -of-particle operator degenerates to
the number $\hat N\equiv N$. To accomplish this, one may resort to the
Girardeau-Arnowitt approach [24,25] introducing the so-called
number-conserving field operators
$$
\al_k = \hat N_0^{-1/2} a_0^\dgr a_k\; , \qquad
\al_k^\dgr = a_k^\dgr a_0 \hat N_0^{-1/2} \; .
$$
This representation, however, is valid only when the number of condensed
particles is large, $N_0\gg 1$, so that it is not applicable in the vicinity
of the condensation point, when $N_0\ra 0$. Also, such a canonical
representation in approximate calculations yields, as is known [24], a gap in
the spectrum. It was noticed by Girardeau [62] and showed by Takano [63] that
to get a self-consistent theory in the canonical ensemble requires to use the
whole Hamiltonian, without reducing it to approximate forms. But the theory
with a whole Hamiltonian for interacting particles has no exact solution. So,
the usage of the canonical ensemble is unpractical in analytic investigations,
though it may be employed for numerical computations [64,65].

Concluding, the choice of an ensemble is, generally speaking, a matter of
convenience. But in any case, the chosen ensemble must be representative,
which necessitates to accurately take into account all conditions uniquely
defining the considered system. The choice of an ensemble is intimately
connected with the choice of the appropriate field variables, which should
not be confused. Employing inappropriate variables, that is, using a
nonrepresentative ensemble may result in the appearance of inconsistences and
paradoxes. For example, resorting to the canonical ensemble, one should use
the Girardeau-Arnowitt representation with the number-conserving field
operators [24,25] or the Carusotto-Castin-Dalibard representation based on
stochastic fields [64,65]. If one prefers the standard grand ensemble, then
the Faddeev-Popov representation is appropriate [61]. A nonstandard variant of
the grand ensemble, due to Hugenholtz and Pines [9], can also be used. But
when the classical Bogolubov representation [1--4] is chosen, where the field
variables of condensed and uncondensed particles are separated from each
other, being independent and orthogonal, then the approach developed in the
present paper is the most convenient, being completely self-consistent.

\vskip 5mm

{\bf Acknowledgement}. I am grateful to the German Research Foundation
for financial support and to the Physics Departments of the Free University
of Berlin and University of Essen for hospitality. I appreciate useful
discussions with M. Girardeau, R. Graham, H. Kleinert, and E. Yukalova.

\newpage

{\Large{\bf Appendix A}}

\vskip 5mm

Here we prove relation (50). From definition (36) for the grand potential
with Hamiltonian (30), it immediately follows that
$$
N_0 = -\; \frac{\prt\Om}{\prt\mu_0} \; , \qquad
N_1 = -\; \frac{\prt\Om}{\prt\mu_1} \; .
$$
On the other hand,
$$
N= - \; \frac{\prt\Om}{\prt\mu} = N_0 + N_1 \; .
$$
Comparing the above equations, we get
$$
\frac{\prt\Om}{\prt\mu} = \frac{\prt\Om}{\prt\mu_0} +
\frac{\prt\Om}{\prt\mu_1} \; .
$$
Differentiating again the latter equation, we have
$$
\frac{\prt^2\Om}{\prt\mu^2} =
\frac{\prt^2\Om}{\prt\mu_0^2} + \frac{\prt^2\Om}{\prt\mu_1^2} +
2\; \frac{\prt^2\Om}{\prt\mu_0\prt\mu_1} \; .
$$
By direct calculations, we find that
$$
\frac{\prt^2\Om}{\prt\mu_0^2} = -\bt\Dlt^2(\hat N_0) \; , \qquad
\frac{\prt^2\Om}{\prt\mu_1^2} = -\bt\Dlt^2(\hat N_1) \; , \qquad
\frac{\prt^2\Om}{\prt\mu_0\prt\mu_1} =
- \bt\; {\rm cov}(\hat N_0,\hat N_1) \; ,
$$
where the covariance of two operators, $\hat A$ and $\hat B$, is
$$
{\rm cov}(\hat A,\hat B) \equiv \frac{1}{2}<\hat A\hat B +
\hat B\hat A> - <\hat A><\hat B> \; .
$$
Summarizing these equations, and using the property of the dispersion
of a composite operator,
$$
\Dlt^2(\hat A+\hat B) = \Dlt^2(\hat A) + \Dlt^2(\hat B) +
2{\rm cov}(\hat A,\hat B) \; ,
$$
we obtain
$$
\frac{\prt N}{\prt\mu} = -\; \frac{\prt^2\Om}{\prt\mu^2} =
\bt\Dlt^2(\hat N) \; ,
$$
which proves relation (50).

\newpage

{\Large{\bf Appendix B}}

\vskip 5mm

The fact that the terms linear in $\psi_1$ or $\psi_1^\dgr$ must be absent
in the Hamiltonian, in order to satisfy constraint (25), can be proved in
two ways.

We may, first, consider a quadratic Hamiltonian approximating the exact one.
Linear terms in such a Hamiltonian can also be present. Quadratic Hamiltonians
of this type can be diagonalized with the help of exact canonical
transformations [45]. Rigorous mathematical properties of the corresponding
transformations, called nonuniform, are expounded in the book by Berezin [42].
After the Hamiltonian is diagonalized, it is straightforward to calculate
explicitly all averages, which show that the linear terms in the Hamiltonian
induce nonzero $<\psi_1>$ and, vice versa, zero linear terms lead to
$<\psi_1>=0$. Then one should consider perturbation theory, starting with
the diagonalized quadratic from, and check that zero linear terms yield
$<\psi_1>=0$ in all orders of the theory. This way is rather cumbersome, and
below another method is presented.

Let us consider the general Hamiltonian (52). Its term, linear in $\psi_1$
and $\psi_1^\dgr$, is
$$
H^{(1)} = \int \psi_1^\dgr(\br) \left ( - \; \frac{\nabla^2}{2m} + U
\right ) \eta(\br)\; d\br +
\int \psi_1(\br) \left ( - \; \frac{\nabla^2}{2m} + U
\right ) \eta^*(\br)\; d\br 
$$
$$
+\; \int \Phi(\br-\br') \left [ 
\psi_1^\dgr(\br)|\eta(\br')|^2 \eta(\br) +
\eta^*(\br)|\eta(\br')|^2 \psi_1(\br) \right ]\; d\br d\br' \; .
$$
The equation of motion (35) can be represented as
$$
i\; \frac{\prt}{\prt t} \; \psi_1(\br,t) =
\frac{\dlt[H - H^{(1)}]}{\dlt\psi_1^\dgr(\br,t)} +
\frac{\dlt H^{(1)}}{\dlt\psi_1^\dgr(\br,t)} \; .
$$
The first term here gives the right-hand side of Eq. (60). The second term
results in the form
$$
\frac{\dlt H^{(1)}}{\dlt\psi_1^\dgr(\br,t)} =
C(\br,t)\; \hat 1_\cF \; ,
$$
proportional to the unit operator $\hat 1_\cF$ in the Fock space
$\cF(\psi_1)$ generated by the field operators $\psi_1^\dgr$, with the
nonoperator complex function
$$
C(\br,t) \equiv \left [ -\; \frac{\nabla^2}{2m} + U +
\int \Phi(\br-\br') |\eta(\br',t)|^2 d\br'\right ] \eta(\br,t) -
\lbd(\br,t) \; .
$$
The equation of motion, written above, is an operator equality defined on
$\cF(\psi_1)$. The operator equality assumes that it holds true at least in
the weak sense implying the equality of all matrix elements for the states
from the space the operators are defined on. In the present case, the latter
means the space $\cF(\psi_1)$. The vacuum state of this space is defined in
the standard way as the vector $|0>$, for which $\psi_1(\br,t)|0>=0$.
Considering for the equation of motion, with respect to $\psi_1$, the matrix
element over the vacuum state, and, taking into account constraint (25), we
get
$$
C(\br,t)= 0 \; .
$$
The latter equation results in the Lagrange multiplier (54), which yields
$H^{(1)}=0$. Thus, in order to preserve constraint (25), the linear terms
in the Hamiltonian must be zero.

The physical meaning of the proved theorem, that $<\psi_1>=0$ necessarily
requires $H^{(1)}=0$, is quite evident. The terms in the Hamiltonian, linear
in $\psi_1$ and $\psi_1^\dgr$ describe the physical processes of annihilation
and creation of single particles. If such processes were permitted, then the
function $C(\br,t)$, introduced above, is not zero. Then the evolution
equation for the average $<\psi_1(\br,t)>$ contains the term
$$
\int_0^t  C(\br,t')\; dt'
$$
generating the nonzero value of $<\psi_1>$.

One should not confuse the considered situation with the often used method
of specially adding to the Hamiltonian the terms, linear in $\psi_1$ and
$\psi_1^\dgr$, in order to break the gauge symmetry. Then, of course, the
average $<\psi_1>$ is not zero. But at the end of calculations, one always
set the additional linear terms to become zero, hence restoring the property
$<\psi_1>=0$. This procedure is what is called the Bogolubov method of
infinitesimal sources [3,4]. These sources are usually lifted after the
thermodynamic limit, but can also be made infinitesimally small in the
process of taking the thermodynamic limit, provided this is done in the
appropriate way [5,66]. In the approach, followed in the present paper, we
do not need to introduce infinitesimal sources, since the gauge symmetry has
already been broken by the Bogolubov shift (16).

\newpage

{\Large{\bf Appendix C}}

\vskip 5mm

The critical temperature $T_c$ for the Bose system with contact interactions
has been obtained, in the HFB approximation, from expansions for the condensate
fraction (149) and superfluid fraction (150). This temperature was found to
coincide with the critical temperature of the ideal Bose gas. In order to
emphasize the correctness of this result, let us consider, first, the more
general case of an arbitrary interaction potential $\Phi(\br)$, provided it
possesses the standard property of symmetry, such that $\Phi(-\br)=\Phi(\br)$,
and diminishes sufficiently fast with increasing $|\br|$.

At the critical temperature $T_c$, when $\rho_0=0$ and $\rho_1=\rho$, Eq.
(112) reduces to
$$
\om_k = \frac{k^2}{2m} + \frac{1}{V}\; \sum_p n_p(\Phi_{k+p}-
\Phi_p ) \; .
$$
The Fourier transform
$$
\Phi_{k+p} = \int \Phi(\br) e^{-i(\bk+\bp)\cdot\br} \; d\br
$$
can be simplfied remembering that, by assumption, the interaction potential
diminishes fast with increasing $r\equiv|\br|$. Then in the above integral,
one can expand $e^{-i\bk\cdot\br}$ in powers of $\bk\cdot\br$ up to the 
second order. As a result, we get
$$
\Phi_{k+p} \cong \left ( 1 \; - \; \frac{1}{6}\; k^2 r_0^2 \right )
\Phi_p \; ,
$$
where the notation for the effective interaction radius $r_0$ is introduced,
defined by the equation
$$
r_0^2 \equiv \frac{\int r^2\Phi(\br)d\br}{\int\Phi(\br)d\br} \; .
$$
In this way, we find
$$
\om_k = \frac{k^2}{2m} \; - \; \frac{1}{6V}\; \sum_p n_p
\Phi_p k^2 r_0^2 \; .
$$
The critical temperature is given by the equation
$$
\rho = \int n_k \; \frac{d\bk}{(2\pi)^3} \; ,
$$
in which
$$
n_k =\left ( e^{\om_k/T_c} - 1\right )^{-1} \; .
$$
From here we obtain
$$
T_c = \frac{2\pi}{m^*} \left [
\frac{\rho}{\zeta(3/2)} \right ]^{2/3} \; ,
$$
where the effective mass is
$$
m^* \equiv
\frac{m}{1-\frac{mr_0^2}{3}\int n_k\Phi_k\frac{d\bk}{(2\pi)^3}} \; .
$$
Thus, for nonlocal interactions, with an interaction radius $r_0$, the
effective mass increases, so that the critical temperature diminishes, as
compared to the critical temperature of the ideal Bose gas.

However, for the contact interaction potential (83), we have $r_0=0$, hence
$m^*=m$, and $T_c$, in the frame of the HFB mean-field approximation,
coincides with the ideal gas condensation temperature.

This conclusion is in agreement with other studies of the critical
temperature for interacting Bose gas. There exists quite a number of such
investigations, as reviewed in Refs. [31,49]. The most accurate of these
calculations are those employing Monte Carlo simulations [67--71] and those
based on the optimized perturbation theory [72--74], as has been done in
Refs. [75--80]. These investigations show that the first correction to the
critical temperature comes from effects beyond the mean-field approximation.


\newpage

\begin{thebibliography}{99}
\bibitem{1}
N.N. Bogolubov, J. Phys. (Moscow) 11 (1947) 23.

\bibitem{2}
N.N. Bogolubov, Moscow Univ. Phys. Bull. 7 (1947) 43.

\bibitem{3}
N.N. Bogolubov, Lectures on Quantum Statistics, Gordon and Breach,
New York, 1967, Vol. 1.

\bibitem{4}
N.N. Bogolubov, Lectures on Quantum Statistics, Gordon and Breach,
New York, 1970, Vol. 2.

\bibitem{5}
V.I. Yukalov, Phys. Rep. 208 (1991) 395.

\bibitem{6}
V.I. Yukalov, Phys. Rev. E 72 (2005) 066119.

\bibitem{7}
P.C. Hohenberg, P.C. Martin, Ann. Phys. (N.Y.) 34 (1965) 291.

\bibitem{8}
J. Ginibre, Commun. Math. Phys. 8 (1968) 26.

\bibitem{9}
N.M. Hugenholtz. D. Pines, Phys. Rev. 116 (1959) 489.

\bibitem{10}
G. Baym, Phys. Rev. 127 (1962) 1391.

\bibitem{11}
J.M. Cornwall, R. Jackiw, E. Tomboulis, Phys. Rev. D 10 (1974) 2428.

\bibitem{12}
V.N. Popov, J. Exp. Theor. Phys. 20 (1965) 1185.

\bibitem{13}
V.N. Popov, Functional Integrals in Quantum Field Theory and
Statistical Physics, Reidel, Dordrecht, 1983.

\bibitem{14}
V.N. Popov, Functional Integrals and Collective Modes, Cambridge
University, New York, 1987.

\bibitem{15}
V.I. Yukalov, E.P. Yukalova, Laser Phys. Lett. 2 (2005) 506.

\bibitem{16}
V.I. Yukalov, Laser Phys. Lett. 2 (2005) 156.

\bibitem{17}
N. Shohno, Prog. Theor. Phys. 31 (1964) 553.

\bibitem{18}
L. Reatto, J.P. Straley, Phys. Rev. 183 (1969) 321.

\bibitem{19}
D.A.W. Hutchinson, K. Burnett, R.J. Dodd, S.A. Morgan, 
M. Rusch, E. Zaremba, N.P. Proukakis, M. Edwards, C.W. Clark, 
J. Phys. B 33 (2000) 3825.

\bibitem{20}
M. Fliesser, J. Reidl, P. Szepfalusy, R. Graham, 
Phys. Rev. A 64 (2001) 013609.

\bibitem{21}
Y.B. Ivanov, F. Riek, J. Knoll, Phys. Rev. D 71 (2005) 105016.

\bibitem{22}
T. Kita, J. Phys. Soc. Jpn. 74 (2005) 1891.

\bibitem{23}
T. Kita, J. Phys. Soc. Jpn. 75 (2006) 044603.

\bibitem{24}
M. Girardeau, R. Armowitt, Phys. Rev. 113 (1959) 755.

\bibitem{25}
M. Girardeau, J. Math. Phys. 3 (1962) 131.

\bibitem{26}
M. Luban, Phys. Rev. 128 (1962) 965.

\bibitem{27}
M. Luban, W.D. Grobman, Phys. Rev. Lett. 17 (1966) 182.

\bibitem{28}
G. Baym, G. Grinstein, Phys. Rev. D 15 (1977) 2897 (1977).

\bibitem{29}
V.I. Yukalov, E.P. Yukalova, Phys. Part. Nucl. 28 (1997) 37.

\bibitem{30}
V.I. Yukalov, E.P. Yukalova, Physica A 243 (1997) 382.

\bibitem{31}
J.O. Andersen, Rev. Mod. Phys. 76 (2004) 599.

\bibitem{32}
V.I. Yukalov, H. Kleinert, Phys. Rev. A  73 (2006) 063612.

\bibitem{33}
V.I. Yukalov, E.P. Yukalova, Phys. Rev. A 74 (2006) 063623.

\bibitem{34}
J.W. Gibbs, Collected Works, Longmans, New York, 1931, Vol. 2.

\bibitem{35}
D. ter Haar, Elements of Statistical Mechanics. Reinhart,
New York, 1954.

\bibitem{36}
D. ter Haar, Rep. Prog. Phys. 24 (1961) 304.

\bibitem{37}
E.T. Janes, Phys. Rev. 106 (1957) 620.

\bibitem{38}
E.T. Janes, Phys. Rev. 108 (1957) 171.

\bibitem{39}
M.D. Girardeau, Phys. Lett. A 30 (1969) 442.

\bibitem{40}
V.I. Yukalov, Statistical Green's Functions, Queen's University,
Kingston, 1998.

\bibitem{41}
H. Kleinert, Path Integrals, World Scientific, Singapore, 2004.

\bibitem{42}
F.A. Berezin, Method of Second Quantization, Academic, New York 1966.

\bibitem{43}
V.I. Yukalov, Laser Phys. 16 (2006) 511.

\bibitem{44}
H. Umegawa, H. Matsumoto, M. Tachiki, Thermo Field Dynamics and
Condensed States, North-Holland, Amsterdam, 1982.

\bibitem{45}
N.N. Bogolubov, N.N. Bogolubov Jr., Introduction to Quantum
Statistical Mechanics, Gordon and Breach, Lausanne, 1994.

\bibitem{46}
L. Pitaevskii, S. Stringari, Bose-Einstein Condensation in Dilute
Gases, Clarendon, Oxford, 2003.

\bibitem{47}
P.W. Courteille, V.S. Bagnato, V.I. Yukalov, Laser Phys. 11 (2001) 
659.

\bibitem{48}
K. Bongs, K. Sengstock, Rep. Prog. Phys. 67 (2004) 907.

\bibitem{49}
V.I. Yukalov, Laser Phys. Lett. 1 (2004) 435.

\bibitem{50}
H.H. S\o rensen, D.V. Fedorov, A.S. Jensen, Am. Inst. Phys. Conf.
Proc. 777 (2005) 12.

\bibitem{51}
A.Z. Patashinsky, V.L. Pokrovsky, Fluctuational Theory of Phase
Transitions, Nauka, Moscow, 1982.

\bibitem{52}
K.G. Wilson, J. Kogut, Phys. Rep. 12 (1974) 75.

\bibitem{53}
F.W. Wirth, R.B. Hallock, Phys. Rev. B 35 (1987) 34.

\bibitem{54}
W. Krauth, Phys. Rev. Let. 77 (1996) 3695.

\bibitem{55}
K. Nho, D. Blume, Phys. Rev. Lett. 95 (2005) 193601.

\bibitem{56}
K. Nho, D.P. Landau, Phys. Rev. A 73 (2006) 033606.

\bibitem{57}
D.M. Ceperley, Rev. Mod. Phys. 67 (1995) 279.

\bibitem{58}
M. Boninsegni, N.V. Prokofiev, B.V. Svistunov, e-print physics/0605225
(2006).

\bibitem{59}
S. Pilati, K. Sakkos, J. Boronat, J. Casulleras, S. Giorgini, e-print
cond-mat/0607721 (2006).

\bibitem{60}
J. Gavoret and P. Nozi\`eres, Ann. Phys. (N.Y.) 28 (1964) 349.

\bibitem{61}
V.N. Popov and L.D. Faddeev, J. Exp. Theor. Phys. 47 (1964) 1315.

\bibitem{62}
M. Girardeau, Phys. Rev. 115 (1959) 1090.

\bibitem{63}
F. Takano, Phys. Rev. 123 (1961) 699.

\bibitem{64}
I. Carusotto, Y. Castin, J. Dalibard, Phys. Rev. A 63 (2001) 023606.

\bibitem{65}
I. Carusotto, Y. Castin, J. Phys. B 34 (2001) 4589.

\bibitem{66}
V.I. Yukalov, Int. J. Mod. Phys. B 5 (1991) 3235.

\bibitem{67}
P. Arnold, G. Moore, Phys. Rev. Lett. 87 (2001) 120401.

\bibitem{68}
P. Arnold, G. Moore, Phys. Rev. E 64 (2001) 066113.

\bibitem{69}
V.A. Kashurnikov, N.V. Prokofiev, B.V. Svistunov, Phys. Rev. Lett.
87 (2001) 120402.

\bibitem{70}
N.V. Prokofiev, B.V. Svistunov, Phys. Rev. Lett. 87 (2001) 160601.

\bibitem{71}
K. Nho, D.P. Landau, Phys. Rev. A 70 (2004) 053614.

\bibitem{72}
V.I. Yukalov, Moscow Univ. Phys. Bull. 31 (1976) 10.

\bibitem{73}
V.I. Yukalov, E.P. Yukalova, Ann. Phys. (N.Y.) 277 (1999) 219.

\bibitem{74}
V.I. Yukalov, E.P. Yukalova, Chaos Solit. Fract. 14 (2002) 839.

\bibitem{75}
B. Kastening, Phys. Rev. A 68 (2003) 061601.

\bibitem{76}
B. Kastening, Laser Phys. 14 (2004) 586.

\bibitem{77}
B. Kastening, Phys. Rev. A 69 (2004) 043613.

\bibitem{78}
J.L. Kneur, A. Neveu, M.B. Pinto, Phys. Rev. A 69 (2004)
053624.

\bibitem{79}
B. Kastening, Phys. Rev. A 70 (2004) 043621.

\bibitem{80}
J.L. Kneur, M.B. Pinto, Phys. Rev. A 71 (2005) 033613.

\end{thebibliography}
\end{document}